\documentclass{jaa}

\usepackage{graphicx}
\usepackage{amsmath}	
\usepackage{amssymb}
\usepackage{url}

\usepackage{natbib}
\begin{document}\sloppy

\title{Recent advances in the 3D kinematic Babcock-Leighton solar dynamo modeling}


\author{Gopal Hazra\textsuperscript{*}}
\affilOne{School of Physics, Trinity College Dublin, Dublin 2, Ireland.\\}


\twocolumn[{

\maketitle

\corres{gopal.hazra@tcd.ie}


\begin{abstract}
 In this review, we explain recent progress made in the Babcock-Leighton dynamo models for the Sun, which have been most successful to explain various properties of the solar cycle. In general, these models are 2D axisymmetric and the mean-field dynamo equations are solved in the meriodional plane of the Sun. Various physical processes (e.g., magnetic buoyancy and Babcock-Leighton mechanism) involved in these models are inherently 3D process and could not be modeled properly in a 2D framework. After pointing out limitations of 2D models (e.g., Mean-field Babcock-Leighton dynamo models and Surface Flux Transport models), we describe recently developed next generation 3D dynamo models that implement more sophisticated flux emergence algorithm of buoyant flux tube rise through the convection zone and capture Babcock-Leighton process more realistically than previous 2D models. The detailed results from these 3D dynamo models including surface flux transport counterpart are presented. We explain the cycle irregularities that are reproduced in 3D dynamo models by introducing scattering around the tilt angle only. Some results by assimilating observed photospheric convective velocity fields into the 3D models are also discussed, pointing out the wide opportunity that these 3D models hold to deliver.
\end{abstract}


}]



\section{Introduction}
Since 1955 after Eugene Parker's first fundamental idea \citep{Parker55a} on origin of solar magnetic cycle, it is almost more than half a century and still we do not understand origin of solar magnetic cycle very well. Understanding solar magnetic cycle is very important not only because it provides us a great opportunity to test our existing theory of plasma physics but also it has an utmost societal importance. The violent solar disturbances (e.g., Solar flares, Coronal Mass Ejections) that are driven by the magnetic field of the Sun, have a strong dependence on the solar magnetic cycle and can affect the space climate tremendously. Also, study of the solar magnetic cycle gives us insights to understand the magnetic cycle of other solar-type stars \citep{KKC14, HJKK19} and its effect on the atmosphere of their exoplanets. \citep{HVD20}.   

Soon after discovery of magnetic field in the sunspot regions \citep{Hale1909}, it was realized that the solar cycle or sunspot cycle is nothing but the magnetic cycle of the Sun. Efforts had been started to understand why solar magnetic field behaves in a particular fashion with a cyclic period of 11-year. The non-linear interaction between turbulent plasma motion and magnetic field inside the Solar Convection Zone (SCZ) is responsible for amplification of magnetic field. This non-linear interaction can be understood by solving a set of MHD equations which govern the behavior of plasma and magnetic field as given below

\begin{equation}
\label{eqnc1:MHD1}
\frac{\partial {\bf v} }{\partial t} + ({\bf v}.\nabla){\bf v}  = {\bf F} - \frac{1}{\rho}\nabla p + \frac{1}{\rho c}{\bf j}\times{\bf B} + \nu \nabla^2{\bf v}
\end{equation}
\begin{equation}
\label{eqnc1:MHD2}
\frac{\partial {\bf B}}{\partial t} = \nabla \times ({\bf v}\times{\bf B}) + \eta\nabla^2{\bf B},
\end{equation}
where ${\bf v}$, ${\bf B}$, $\nu$ and $\eta$  are the velocity, magnetic field, viscosity and magnetic diffusivity respectively. $F$ represents the gravity force. Other terms are as usual. Where as the fundamental equations are well established, one of the major challenges to develop a solar dynamo model is to handle the turbulent convective motions properly inside th SCZ. The turbulent stresses in the convection zone drive the large-scale plasma flows such as differential rotation and meridional circulation which are very crucial for operation of solar dynamo. Hence modeling turbulence in a proper way is very important to understand large-scale flows and solar dynamo theory.

Historically the mean field approach of turbulence played a major role in development of the dynamo theory. In the mean field approach, velocity field and magnetic field are split into two parts, mean and fluctuating parts:
\begin{equation}\label{eq:fluc}
  \bf v = \bar{v} + v', \bf B = \bar{B} + \bf{B}'
\end{equation}
where overline indicates the mean quantities and prime denotes the fluctuation from the mean. By substituting equation {\ref{eq:fluc}} in the magnetic induction equation {\ref{eqnc1:MHD2}}, we get
\begin{equation}
\frac{\partial {\bf B}}{\partial t} = \nabla \times ({\bf v}\times{\bf B}) + \nabla \times \mathbf{\xi} + \eta\nabla^2{\bf B}
\end{equation}
where ${\mathbf {\xi}}$ = $\overline{{\bf v'}\times{\bf B'}}$ is the mean electromotive force (EMF) that sustains the dynamo action in the Sun. For homogeneous isotropic turbulence, mean EMF can be written as
\begin{equation}
  {\mathbf \xi} = \alpha \bar{\bf B} - \beta\nabla \times \bar{\bf B}.
\end{equation}
Here $\alpha$ represents the classical helical $\alpha$-effect and $\beta$ represents the turbulent diffusivity (see \citet{Chou98} for details).


At present, the most promising framework to explain the properties of solar magnetic field is the  Babcock-Leighton (BL)/Flux Transport dynamo (FTD) model \citep{CSD95, Durney95, Durney97, DC99, CNC04,HKC14,Karak14}. These models are a class of mean field models where the Babcock-Leighton $\alpha$ effect is considered instead of the classical helical $\alpha$ effect, and meridional flow plays a very important role. The mean solar magnetic field is generally assumed to be axisymmetric in these models and can be decomposed into two parts, namely the toroidal and poloidal components. Parker (1955) first suggested that the solar magnetic cycle is a result of oscillation between toroidal field and poloidal field, and the toroidal and poloidal fields sustain each other through a cyclic feedback process. For the Sun, as equator rotates faster than pole, the differential rotation stretches the poloidal field and generates the toroidal field. When toroidal field becomes magnetically buoyant, it rises up and pierces the surface to create the sunspots. The bipolar sunspots always have an angle in between them (tilt angle) with respect to the equatorial line because of the Coriolis force that acted on the toroidal field while it rises through the convection zone due to magnetic buoyancy \citep{DSilva93}. Since the Coriolis force increases with increasing latitudes, the tilt angle also increases as sunspots erupt at the higher latitudes, which is first observed by Joy and known as Joy's law \citep{Hale19}. Also, sunspots are the regions of strong magnetic field and they diffuse. As a result, the leading polarity sunspots that are near to the equator cancel with opposite polarity sunspots from the opposite hemisphere. The trailing polarity sunspots from each hemisphere advect to the polar region and generate the large scale poloidal field. This whole mechanism is called Babcock-Leighton mechanism \citep{Bab61, Leighton69}. This mechanism plays a very crucial role in the BL dynamo model by converting the toroidal field to the poloidal field. Once large scale poloidal field is generated on the surface of the Sun, it is advected to the bottom of the convection zone by meridional circulation or the turbulent diffusion, depending upon which one has the faster time scale. As the BL mechanism needs involvement of the longitudinal co-ordinate over the surface of the Sun and most of the BL/FTD model follows a 2D axisymmetric formulation in which the magnetic induction equation is solved in the meridional plane of the Sun, a parametric approach has been widely used to capture the BL mechanism in 2D BL models.

The BL mechanism is observationally very well supported. \citet{KO11a} calculated the global poloidal field by multiplying tilt angle and the magnetic field strength of active regions for an individual cycle, which is correlated with the strength at minimum of that following cycle. \citet{Dasiespuig10} also found a significant correlation between product of the cycle's averaged tilt angle and the strength of the same cycle with the strength of the next cycle supporting BL mechanism.

There are another class of models that treat the BL mechanism more realistically than the parametric approach used in the 2D axisymmetric BL dynamo models. These are called Surface Flux Transport (SFT) models \citep{WNS89,WNS89a, Baumann04,Jiang14}. Note that these are not dynamo models rather they only consider the evolution of the radial field on the surface of the Sun. In these models, sunspots are directly incorporated and the decay of sunspots due to turbulent diffusivity and corresponding advection of fields to the pole by meridional circulation are modeled by solving radial part of magnetic induction equation on the surface (latitude -longitude plane) of the Sun. They capture realism of the BL mechanism in great detail but they have their own limitations.

In both type of the models (BL dynamo models and SFT models), mean flows (e.g., meridional flow and differential rotation) play a very important role. We have overwhelming amount of data from helioseismology for meanflows \citep{Thompson96, ABC08}. For SFT models, the required surface information of the mean flows is well constrained. But for BL dynamo model, we need information about the mean flows inside whole convection zone, in which the differential rotation is well mapped \citep{Schou98,ABC98} but the exact nature of meridional circulation is still an active field of research. In most of the BL dynamo model, a single cell meridional circulation encompassing whole convection zone with a poleward flow near the surface and an equatorward return flow near the base of convection zone is assumed. The poleward flow near surface is observed by helioseismolgy but detecting the equatorward return flow is an extremely difficult task because of very high noise in the helioseismology data near the bottom of the convection zone. However, recently \citet{Gizon2020} finds a single cell meridional circulation with an equatorward return flow in each hemisphere of the Sun. Also, some of the numerical simulations find the equatorward return flow due to angular momentum balance with a solar like differential rotation \citep{Passos15}.

Most of 2D BL dynamo models are successful in explaining various properties and irregularities of the solar magnetic cycle. But some of the processes involved in the model are observationally not well constrained. The toroidal field generation mechanism from poloidal field by differential rotation is well constrained from helioseismology. But the flux emergence due to magnetic buoyancy and creation of sunspots -- this whole process is inherently 3D process and could not be modeled properly in 2D. Also, due to lack of azimuthal information in 2D models, the realism of BL process can not be captured as it is done in SFT models. But the SFT models have their own limitation for not considering subsurface processes (e.g., subduction of magnetic field by meridional circulation in the polar regions) and 3D vectorial nature of magnetic field. Therefore, the development of 3D dynamo models will help in capturing the 3D processes involved in the dynamo model more realistically and build a bridge between 2D BL dynamo models and SFT models. Also, it will help us with new opportunities to assimilate the observed photospheric data to probe the interior of the solar convection zone.


This review is structured as follows. In the next section, we will briefly describe advantages and disadvantages of the 2D models including SFT models. The formulation of next generation 3D dynamo models based on newly developed flux emergence algorithms and some results are given in Section~3. The advantage of 3D models compared to the 2D models in the light of build up of polar field is discussed in Section~4. In Section~5, we discussed how irregular properties of the solar cycle can be studied by including more  realistic treatment of tilt angle scatter around Joy's law. The opportunity of observed data assimilation in the 3D models and some enlightening results including those data are presented in Section~6. Finally in Section~7, we summarize and conclude all results from 3D models indicating the tremendous possibility that these models have to emerge as the next generation dynamo models.

\begin{figure}[t!]
\centering{\includegraphics[width=.35\textheight]{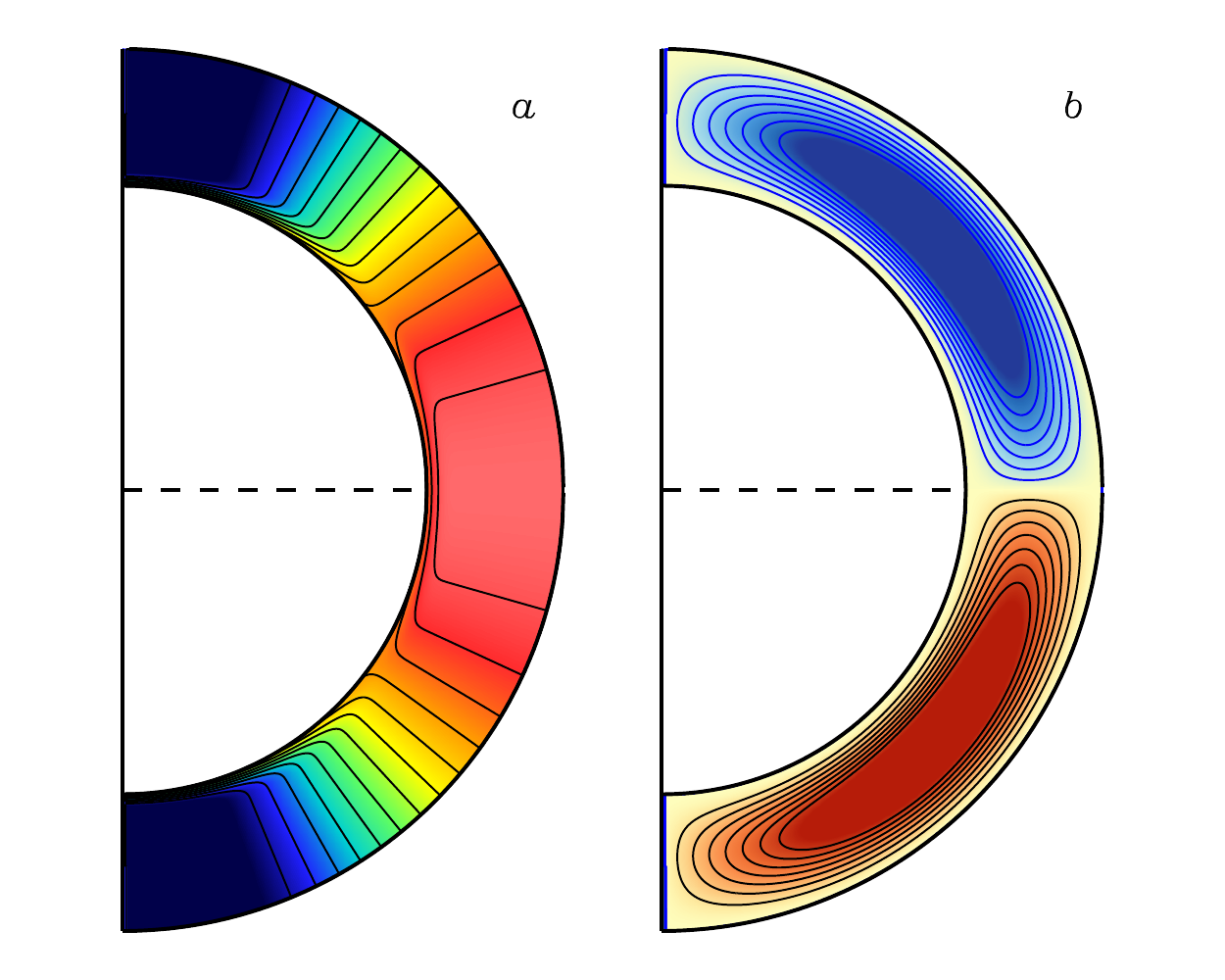}}
\caption{(a) Differential rotation profile. Color scale represents differential rotation with value 350-480 nHz from blue to red color and (b) The meridional flow streamlines. Blue contours show the poleward flow at the surface and an equatorward
flow at the bottom of the convection zone in northern hemisphere
 and red contours show the same at southern hemisphere. From \citet{HCM17}}
 \label{fig:mflows}
\end{figure}

\section{2D Models}\label{sec:2dmodel}
\subsection{2D Axisymmetric BL/FTD models}
The mean axisymmetric magnetic field of the Sun can be written as

\begin{equation}
{\bf B} = B_\phi \hat{\phi} + \nabla \times A{\hat{\phi} },
\end{equation}

where $B_\phi$ is the toroidal field and A$\hat{\phi}$ is the magnetic vector potential, curl of which gives rise to the poloidal field.  The toroidal and poloidal field evolution equations are given below:

\begin{equation}
\label{eq:Aeq}
\frac{\partial A}{\partial t} + \frac{1}{s}({\bf v}.\nabla)(s A)
= \eta_{p} \left( \nabla^2 - \frac{1}{s^2} \right) A + S(r, \theta, t),
\end{equation}
\begin{eqnarray}
\label{eq:Beq}
\frac{\partial B}{\partial t}
+ \frac{1}{r} \left[ \frac{\partial}{\partial r}
(r v_r B) + \frac{\partial}{\partial \theta}(v_{\theta} B) \right]
= \eta_{t} \left( \nabla^2 - \frac{1}{s^2} \right) B \nonumber \\
 + s({\bf B}_p.{\bf \nabla})\Omega + \frac{1}{r}\frac{d\eta_t}{dr}\frac{\partial{(rB)}}{\partial{r}}
\end{eqnarray}

Here $B_p$ is the poloidal field, $s = r\sin\theta$ and other terms are as in usual notation. $S(r,\theta)$ is the source function that incorporates the flux emergence through the convection zone and subsequent BL process.

In most of the dynamo models, observationally motivated various analytical profiles of the differential rotation ($\Omega$) and meridional flow ($v_r,v_\theta$) are used, which are very close to the helioseismology findings. A particular profile of differential rotation and meridional flow is shown in Figure~\ref{fig:mflows}. In general, a single cell meridional circulation is used. Although the equatorward return flow near the bottom of the convection zone for the single cell meridional circulation is extremely difficult to observe by helioseismology, it is needed in order to full fill the mass conservation. However, recently \citet{Gizon2020} found an equatorward flow near the bottom of the convection zone supporting a single cell meridional circulation. This equatorward flow plays a very important role in advecting the toroidal field towards equator against the poleward dynamo wave \citep{Yoshimura75} explaining the equatorward migration of the sunspots. Even in case of the multi-cell meridional circulations inside solar convection zone, this equatorward return flow near the bottom of the CZ is important for dynamo to work \citep{HKC14}.

The parametric approach of modeling magnetic buoyancy and BL process widely varies across different 2D dynamo models. It can be classified in two specific approaches, one local buoyancy and another one as non-local buoyancy. In the local buoyancy treatment, the toroidal field is depleted from the bottom of the CZ once it is more than a critical value and placed on the surface to account for the poloidal field generation. The depleted toroidal field is usually multiplied by an $\alpha$-parameter which is confined near the surface layers. In non-local buoyancy treatment, the toroidal field at the bottom of the convection zone is directly multiplied with the $\alpha$-parameter on the surface to generate poloidal field. For a detail discussion about the treatment of magnetic buoyancy please see \citet{CH16}. In general, both of the treatments of the magnetic buoyancy  reproduce the basic features (e.g.,11-year periodicity, equatorward migration of sunspots, polarity reversal) of observed solar cycle quite well. But depending upon how we treat the magnetic buoyancy in 2D models, many irregularities of the solar cycle (e.g., Waldmeier effect, Correlation of decay rate with next cycle amplitudes) may or may not be reproduced. Also, \citet{Munoz10} pointed out that $\alpha$-parameterization does not correctly depict the relation between the speed of surface meridional flow and strength of the polar field, rather another formalism called Durney's double ring algorithm \citep{Durney95, Durney97} catches the intuitive process of BL mechanism more physically. Given that many parametric formalism of magnetic buoyancy and BL process in 2D models lead to varied results,  next step would be to model the flux emergence due to magnetic buoyancy and subsequent sunspots decay due to BL process using a 3D framework \citep{YM13, MD14, MT16,HCM17, KM18, HM18}

\subsection{SFT models}
The visible part of the BL process on the surface i.e., the dispersion and migration of the sunspot fields after sunspots emerge is well captured in the SFT model. However, unlike BL dynamo models, It only solves the radial component of the magnetic induction equation on the surface of the Sun in the latitude-longitude plane. The radial component of magnetic induction equation on the surface of the Sun (at $r = R_\odot$) is given below

\begin{eqnarray}
\frac{\partial B_r}{\partial t}  = -\frac{1}{R_\odot\sin\theta}\frac{\partial}{\partial \phi}(uB_r) -  \frac{1}{R_\odot\sin\theta}\frac{\partial}{\partial \theta}(vB_r\sin\theta) \nonumber \\
 + \eta_H\left[\frac{1}{R_\odot^2\sin\theta}\frac{\partial}{\partial \theta}\left(\sin\theta\frac{\partial B_r}{\partial \theta}\right) + \frac{1}{R_\odot^2\sin^2\theta}\frac{\partial^2B_r}{\partial \phi^2}\right]
\nonumber\\
+ S(\theta,\phi,t) + D(B_r).
\end{eqnarray}

The radial component of magnetic field ($B_r$) has been used as a passive scaler here that can be mixed and advected to the pole under the effective action of differential rotation i.e., the  velocity in the longitudinal direction (u), latitudinal meridional flow (v)  and turbulent diffusion ($\eta_H$). $S(\theta,\phi,t)$ incorporates the new fluxes emerge from the surface below. $D(B_r)$ is the term that take cares of the decay of magnetic field due to radial diffusion. Historically, this model plays a tremendous role in understanding the BL process and subsequent build up of polar field. In this model, one can study in detail that how individual sunspot pair contributes in the build up of the polar field, and how the latitudinal position of the sunspots and their tilt angle distribution are going to affect the strength of the polar field. The main limitation of the model is not accounting many important physics by ignoring the vectorial nature of the magnetic field and by not incorporating the sub surface processes.  There are some studies that show that the subduction of the poloidal field by the meridional flow sinking underneath the solar surface plays a very important role in the dynamics of the magnetic field \citep{DC94, DC95, CD99}.  Since these processes can not be incorporated in 2D SFT models, the advected radial magnetic flux near the polar region tends to get piled up and it can only be neutralized  by opposite polarity flux advected  there. Therefore, if additional flux of opposite polarity is not advected to  the polar regions, the polar field may reach to an asymptotic value \citep[See figure 6 of][]{Jiang14}.  One may get a secular drift of the polar field while modeling several cycles if the flux  of the succeeding cycle is unable to properly neutralize the polar flux of the preceding cycle. \citet{Baumann04,Baumann06} proposed a way of fixing this problem by adding an ad-hoc decay term corresponding to the radial diffusion which is not included in the SFT model. Hence although, SFT models played very important role in elucidating BL process, it has some inherent limitation that it can not handle the dynamics of magnetic field in the polar regions appropriately.

The next step would be to develop the 3D kinematic BL dynamo models where the fluid motions are still provided and the evolution of the magnetic field would be in 3D. These models can incorporate the attractive features of the 2D BL dynamo models and SFT models, while being free from the limitations of the both of the models.


\section{3D kinematic dynamo models}
3D dynamo models are the next generation dynamo models that implement BL process with high observational fidelity and treat magnetic flux emergence through the SCZ much more realistically than 2D dynamo models. In these models, the total non-axisymmetric magnetic field of the Sun is considered and their evolution is studied by solving the magnetic induction equation in a 3D rotating spherical shell with radius  ranges from $r = 0.69 R_\odot$ to $r = R_\odot$ as given below:

\begin{equation}\label{eq:induction}
\frac{\partial {\bf B}}{\partial t} = \nabla (\times {\bf v} \times {\bf B} - \eta \nabla \times {\bf B}) + S(r,\theta, B,t)
\end{equation}

where ${\bf B}(r,\theta,\phi,t)$ is written in terms of toroidal and poloidal magnetic potentials A and C such that  $${\bf B} = \nabla \times(A {\hat r}) + \nabla \times \nabla \times (C \hat{r}). $$ $\eta$ is the magnetic diffusivity inside solar convection zone and $v$ is the mean flows. In most of the cases, a radial field at the surface and a conducting lower boundary have been used as boundary conditions for solving equation~\ref{eq:induction}. Although the magnetic field is in 3D, the velocity fields are still axisymmetric in general. However \citet{HM18} considered the effect of non-axisymmetric velocity fields to study the BL process.  The  source term $S(r,\theta,t)$ incorporates the BL process and  magnetic flux emergence through the solar convection zone. The inherent 3D non-axisymmetric features of  flux emergence due to magnetic buoyancy is now modeled more realistically in 3D models. Different treatments on every aspects of flux emergence and Babcock-Leighton processes in the 3D framework are discussed in the next subsections.

\subsection{Flux emergence}\label{sec:buoyancy}
\begin{figure*}[t!]
\centering{\includegraphics[width=.75\textheight]{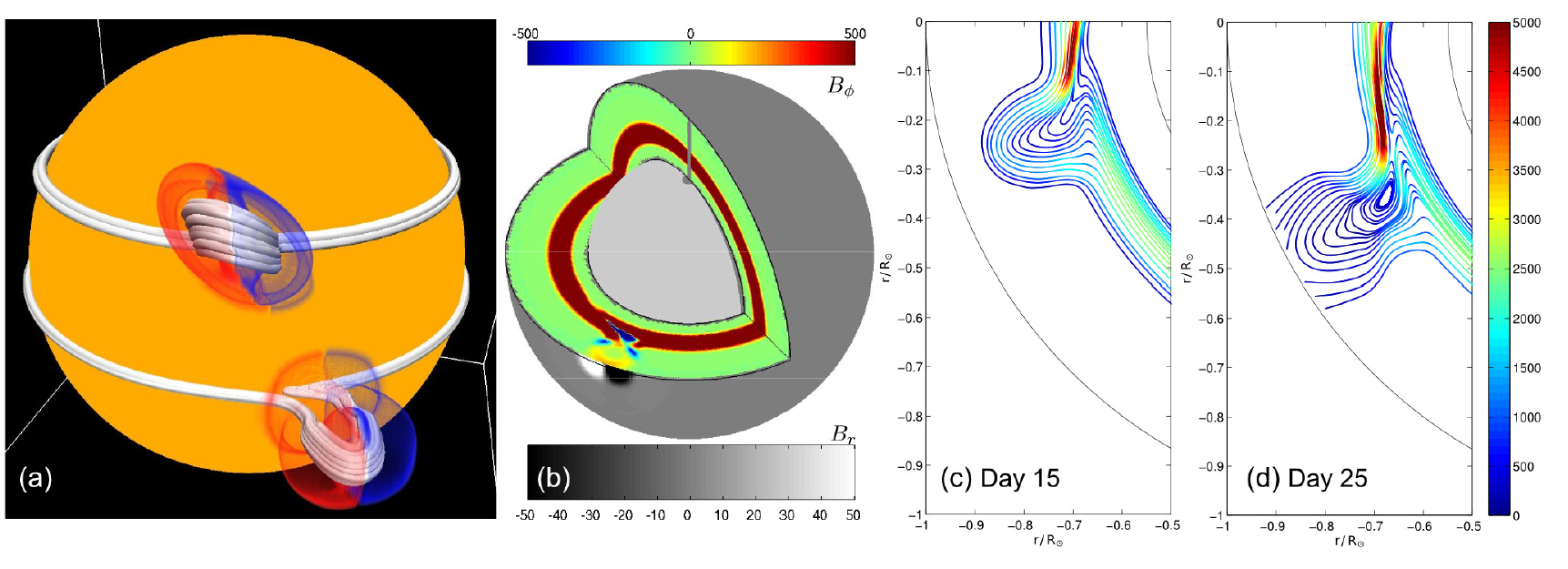}}
\caption{This figure is taken from Yeates and Munoz-Jaramillo (2013) showing a single flux tube emergence on day 25 at two different latitudes (a) at 30$^{\circ}$ N and  0$^{\circ}$ latitudes. Red and blue colors on the surface show positive and negative polarity $B_r$ respectively. (b) Colored contours show a cut of $B_{\phi}$ inside the convection zone and contours on the surface show $B_r$ (on day 25). The magnetic field lines in the equatorial plane at on days 15 and 25 are shown in panel (c) and (d). All colorbars are in units of Gauss.}
\label{fig:f_emerg}
\end{figure*}

First time in a 3D framework, \citet{YM13} modeled full process of three-dimensional emergence of flux tube considering its interaction with convective flows while rising through the solar convection zone. Their procedure is really unique in a way that it incorporates key features of emerging flux tubes, as suggested by the thin-flux tube and anelastic MHD simulations, and allows the flux emergence in a more consistent way than artificial flux deposition on the surface of the Sun. This treatment of flux emergence in the dynamo framework would enhance our understanding of the emergence and decay of sunspots as a source for creating poloidal field from the toroidal field. In figure~\ref{fig:f_emerg}, the emergence of two isolated flux tubes at two different latitudes (0$^{\circ}$ and 30$^{\circ}$) are shown. It is clear from the simulation that the rotational shear of the emerging flux tube leads to relative movement of the flux tube with respect to its roots, which is very important for the magnetic configuration near the eruption site.

Another method that has been developed to incorporate flux emergence and corresponding creation of sunspots is called the 'Spotmaker' algorithm \citep{MD14, MT16, HCM17, KM18}. This method is different than the flux emergence procedure adopted in \citet{YM13}. In Spotmaker algorithm, the spots are placed on the surface of the Sun based on dynamo generated toroidal field near base of the convection zone. In this method, the time required for flux to travel through the convection zone is neglected with respect to time scale of the solar cycle.

\begin{figure*}[htbp!]
\centering{\includegraphics[width=.65\textheight]{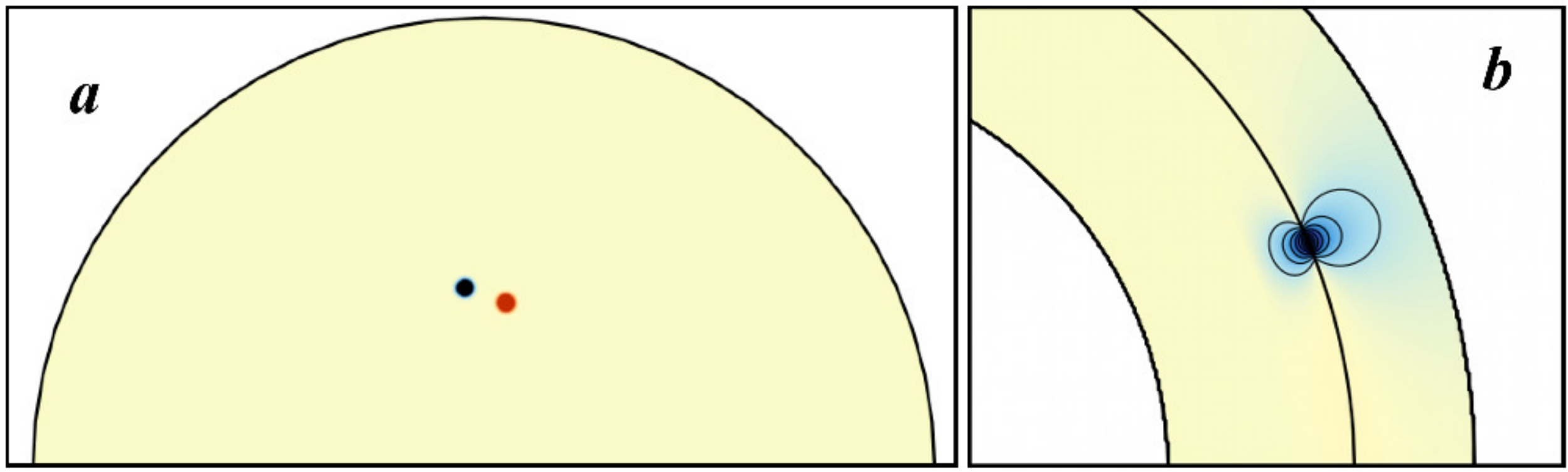}}
\caption{Structure of a typical spot pair produced by the `Spotmaker' algorithm following a Joy's law tilt. (a) Orthogonal projection of $B_r$ at the solar surface associated with two spots at a mid latitude. Red and blue shows positive and negative polarity of sunspots. (b) The poloidal field associated with the spotpairs. The potential field used for subsurface structure extended upto 0.95R$_\odot$. This figure is taken from \citet{MD14}.}
\label{fig:spot}
\end{figure*}

\begin{figure}[htbp!]
\centering{\includegraphics[width=.35\textheight]{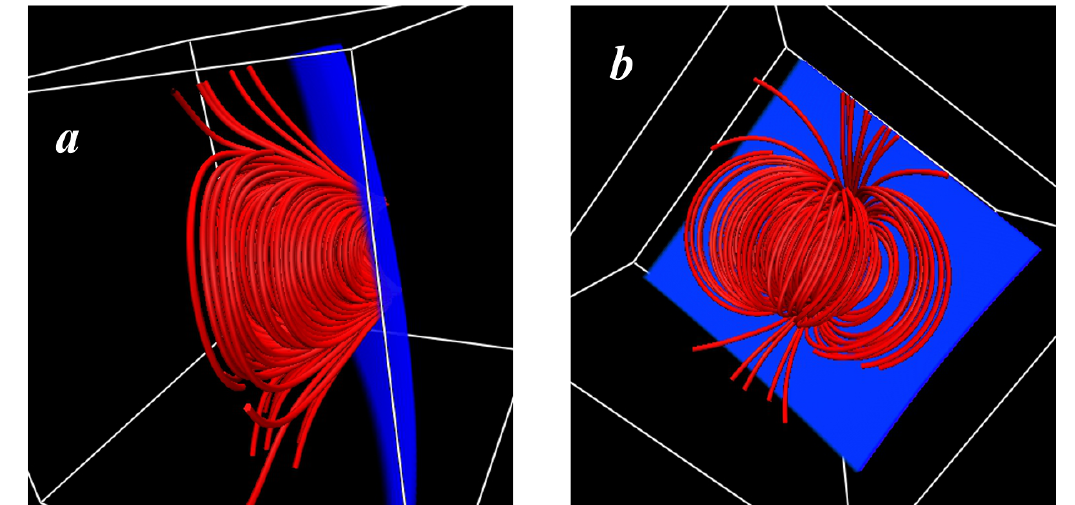}}
\caption{The detailed subsurface structure of magnetic field lines in a sunspot pair is shown from \citet{MT16}. The volume rendering shows magnetic field lines (red contours) below the solar surface at two different vantage points (a) east of the sunspot pair looking west and (b) underneath the sunspot pair looking up. The surface of the sun is shown by the blue surface.}
\label{fig:subsurface}
\end{figure}

As a first step, the spot producing toroidal field is calculated  at the tachocline by averaging toroidal field over the radius from r = $0.70 R_\odot$ to r = $0.71 R_\odot$. Then the bipolar spot is placed once the averaged toroidal field near the tachocline exceeds a threshold value $B_t$ as shown in figure~\ref{fig:spot}(a). The corresponding potential field approximation of the placed sunspot is also shown in figure~\ref{fig:spot}(b). The placement of a spot pair on the surface in latitude and longitude is decided by the location where the averaged toroidal field crosses the threshold value. However after emergence, the spots are not connected to its parent flux tubes. A potential field extrapolation below the surface of each spot pair is used for the subsurface structure. The subsurface structure of each spot is shown in figure~\ref{fig:subsurface}(a) and (b).

After deciding the locations of the spot pair, the timing for sunspots appearance is determined by a time delay probability density function motivated from observed sunspots data. For example, if a sunspot pair appears at time $t_0$, then the timing of the next emergence event will be at $t_1 = t_0 +  \Delta_n$, where $\Delta_n$ is chosen randomly based on the time delay probability distribution function $P(\Delta)$ \citep{MT16}. Once the sunspots pairs are placed on the surface of the sun based on the locations and timing determined by the toroidal field and time delay pdf, their subsequent evolution due to differential rotation, meridional circulation and turbulent diffusion generates the poloidal field naturally via Babcock-Leighton mechanism. The Spotmaker algorithm captures much better the sunspots properties after emergence i.e., the late phase of the flux emergence on the surface while the procedure adopted in the Yeates and Munoz-Jarameillo 2013 captures the early phase of the flux emergence better.

Recently \citet{Kumar19} has employed the dynamical flux emergence by considering upward vortical flows and subsequent evolution of the spots to create the poloidal field. Unlike \citet{YM13}, they are able to obtain a self excited dynamo but this dynamic flux emergence algorithm gives rise to the overlapping sunspot distribution near the minima. A comparative study of different flux emergence algorithm to explain variuos irregular properties might be very helpful to constrain the exact flux emergence method.

\subsection{Dynamo quenching and B-L process}
One of the main issues related to the kinematic approach of dynamo modeling is not accounting for the Lorentz force feedback on the mean flows. Presumably for the Sun, kinematic approach is not at all bad approximation because the observed torsional oscillation i.e., the cyclic variation of the differential rotation is not very significant \citep{AB00,CCC09} and the results obtained from these models are quite in good agreement with the observations. However, in kinematic framework the dynamo needs to be quenched for a given velocity field to suppress its unlimited growth. In the spotmaker algorithm,  the flux being deposited on spot pairs is suppressed by a quenching factor:

\begin{equation}
\Phi_s =  2\Phi_0 {\frac{|{{B}(\theta_s,\phi_s, t)|}}{B_q}}\frac{10^{23}}{1 + \left(\frac{B(\theta,\phi)}{B_q^2}\right)^2} \rm{Mx}
\label{eq:quench1}
\end{equation}
where $B$ is the toroidal field averaged over the tachocline and $B_q$ is the quenching field strength usually assumed 10$^5$ G. The $\Phi_0$ factor is the amplification factor which can be adjusted to make dynamo action to be supercritical. If we choose $\phi_0 \sim 1$, that will give a flux of 10$^{23}$ \rm{Mx} in the strongest active region closed to the observations with the subsurface toroidal field strength equivalent to the quenching field strength.

Another very physical way to incorporate dynamo quenching is by introducing a quenching factor in the tilt angle between the bipolar magnetic regions. As the strongest flux tubes get less affected by the Coriolis force while rising fast through the solar convection zone, we expect the tilt angle to be quenched when cycle strength is strong. \citet{KM17} used the tilt angle quenching as the dynamo saturation mechanism and were able to get a  self-sustained dynamo solution.

\begin{figure*}[htbp!]
\centering\includegraphics[width=.73\textheight]{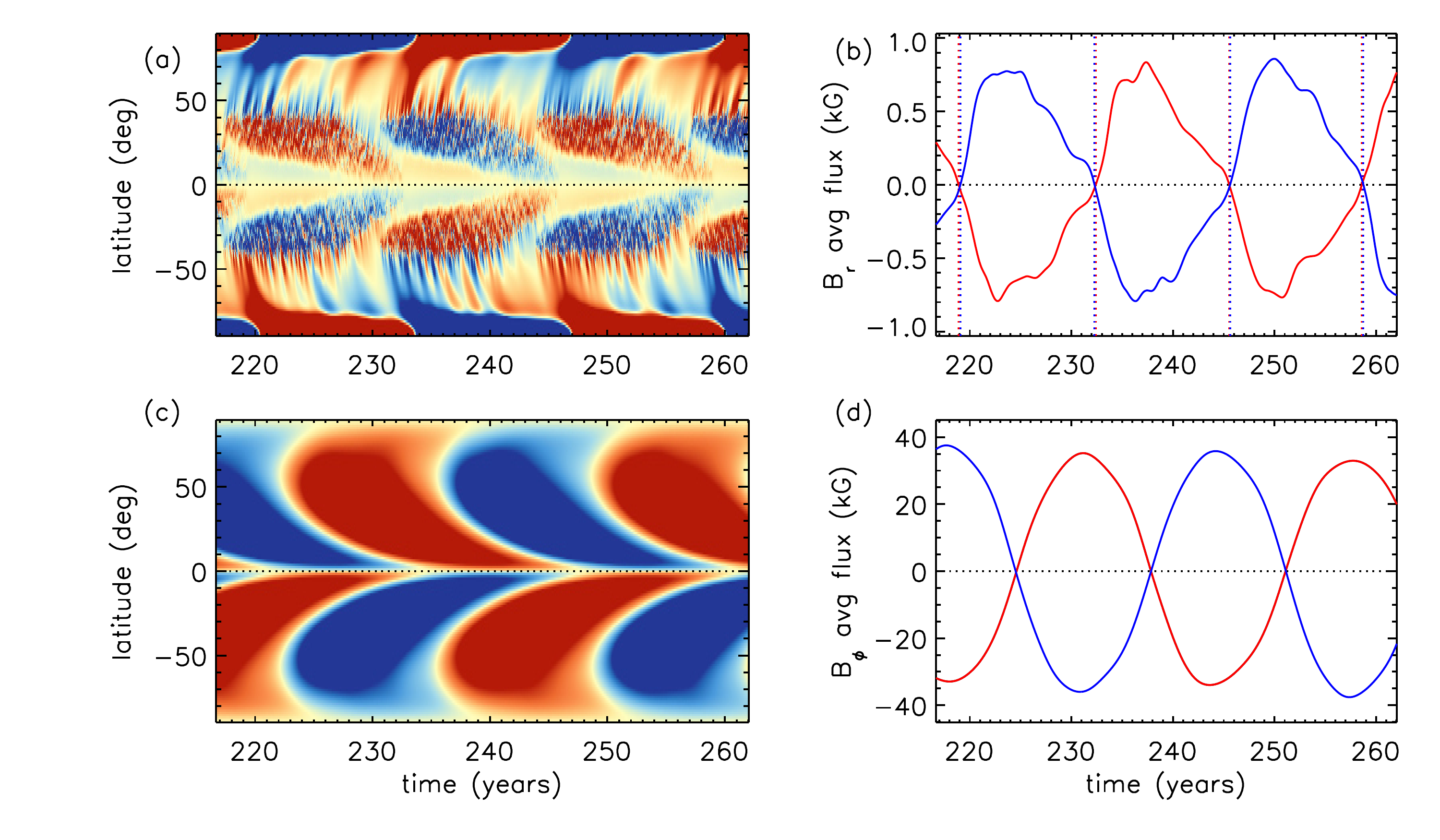}
\caption{Magnetic cycles from a standard STABLE simulation taken from \citet{HM18}. (a) The azimuthal averaged $B_r$ is shown as a function of time and latitudes, highlighting three cycles. Red and blue colors show positive and negative polarity respectively. The color scale is set from −200 G (blue) to +200 G (red). (b) Mean polar field i.e., the averaged radial field over latitudes poleward of $\pm$ 88$^{\circ}$ for the same three cycles are shown.  Polar field reversals are shown using the dotted lines. Blue and red colors correspond to the northern and southern hemispheres respectively. (c) Time–latitude plot of the azimuthal averaged mean toroidal field $\hat{B_\phi}$ at r = 0.7R$_\odot$ (bottom of the convection zone). The color scale saturates at $\pm$ 50 kG, with red and blue denoting eastward and westward field respectively. (d) Evolution of mean toroidal flux near the base of the CZ, averaged over the northern (blue) and southern (red) hemispheres for the same three cycles.}
\label{fig:bfly}
\end{figure*}

\subsection{Evolution of dynamo generated fields}
We present results from the self-sustained 3D kinematic dynamo models called Surface flux Transport And Babcock-Leighton (STABLE) dynamo model here, which have mostly been presented in \citep{MD14, MT16, HCM17, KM17, HM18}. The butterfly diagram is mostly considered as the signature of the cyclic properties of the solar cycle. In figure~\ref{fig:bfly}(a) and \ref{fig:bfly}(c), the butterfly diagram i.e., the time latitude diagram of azimuthal averaged toroidal and radial fields are shown at $r$ = 0.71R$_\odot$ and $r$ = R$_\odot$ respectively. The butterfly diagrams show a cyclic behavior in both toroidal and radial fields with nearly 13-year periodicity. The fine tunning of meridional flow speed or turbulent diffusion can  give us exactly 11-year period of the solar cycle but we want to explore the overall properties of the solar magnetic activity. The equatorward propagation of sunspot field are clear from the radial field evolution diagram (figure~\ref{fig:bfly}(a)). The sunspot producing field i.e., the toroidal field also shows an equator migration with time (figure~\ref{fig:bfly}(c)) presumably due to the equatorward meridional circulation near the bottom of the convection zone. Since there is no physical sunspot number in most of the previous 2D models, the toroidal field at the bottom of the convection zone is considered as a proxy of the sunspot number.  But for 3D model, we have physical sunspots on the surface that contributes to the polar field. The disintegration and migration of sunspot field due to meridional circulation, differential rotation and turbulent diffusion give rise to a poleward migration of trailing flux that reverses the polar field according to the BL mechanism (see figure~\ref{fig:bfly}(a)).

In figures~\ref{fig:bfly}(b) and (d), the evolution of polar flux and averaged toroidal flux over each hemispheres are shown. Polar flux is calculated by averaging radial field on the polar region above $\pm$ 85$^{\circ}$ latitudes, while toroidal flux is calculated over whole convection zone in each hemisphere. The opposite phase difference between evolution of toroidal flux and polar flux makes it clear that polar field reaches a minimum value while toroidal field is maximum i.e., during the solar maximum in accordance with the observation. The 3D dynamo models are capable of reproducing most of the aspect of solar magnetic field with a very good promise to include many observational data for greater understanding of solar magnetic activity as we explain in the next sections.

\begin{figure*}[htbp!]
\centering{\includegraphics[width=.72\textheight]{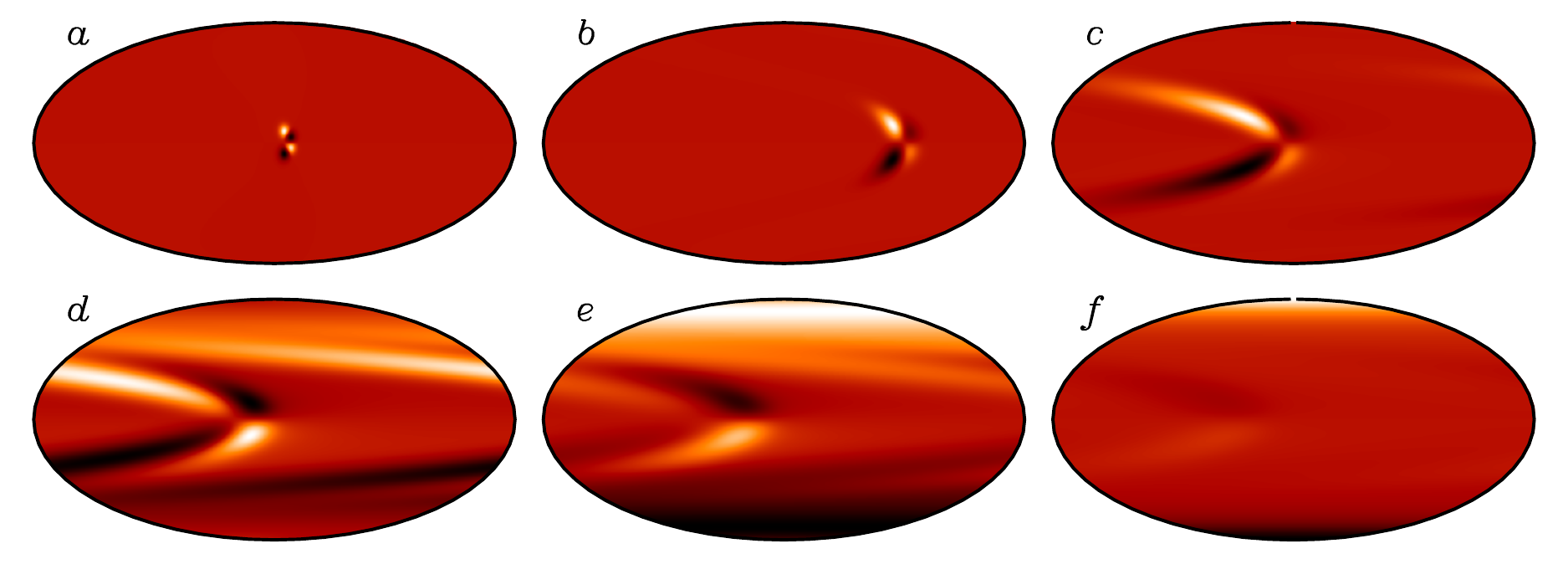}}
\caption{The evolution of radial magnetic fields on the surface of the Sun for sunspot emergence in two hemispheres at $\pm$ 5$^{\circ}$ latitudes during (a) 0.025 yr, (b) 0.25 yr, (c) 1.02 yr, (d) 2.03 yr, (e) 3.05 yr, and (f) 4.06 yr. Here, white and black color show the outward and inward going radial field respectively. In each case, the color scale is set at $\pm$ maximum values of the magnetic field. For example, in case (a), the color scale is set at $\pm$ 4.66 G and  $\pm$ 0.10 G is the color scale for (f). From \citet{HCM17}.}
\label{fig:twopairs}
\end{figure*}
\section{Behavior of surface flux transport in the 3D dynamo model}
The addition of extra azimuthal dimension in the 3D models allows us to investigate the behavior of flux transport on surface of the Sun. Hence one of the main aspects that we can address after constructing a self-excited 3D dynamo models is how individual sunspot pair contributes to the building up of the polar field and whether our understanding gained from the 3D models necessitates the revision of some insights gained from 2D SFT models. In order to do that an individual sunspot pair on each hemisphere is placed at a particular latitude and let them evolve under the axisymmetric mean flows and turbulent diffusion. \citet{HCM17} explored a few cases by putting a single sunspot pair in northern hemisphere, two sunspot pairs symmetrically in two hemispheres and two sunspot pairs in two hemispheres at two different longitudes as well (not symmetric). We discussed here only one case in details, where two sunspot pairs are placed symmetrically in two hemispheres.

\subsection{Build up of polar field from two sunspot pairs in two hemispheres}
\begin{figure*}[htbp!]
\centering{\includegraphics[width=.75\textheight]{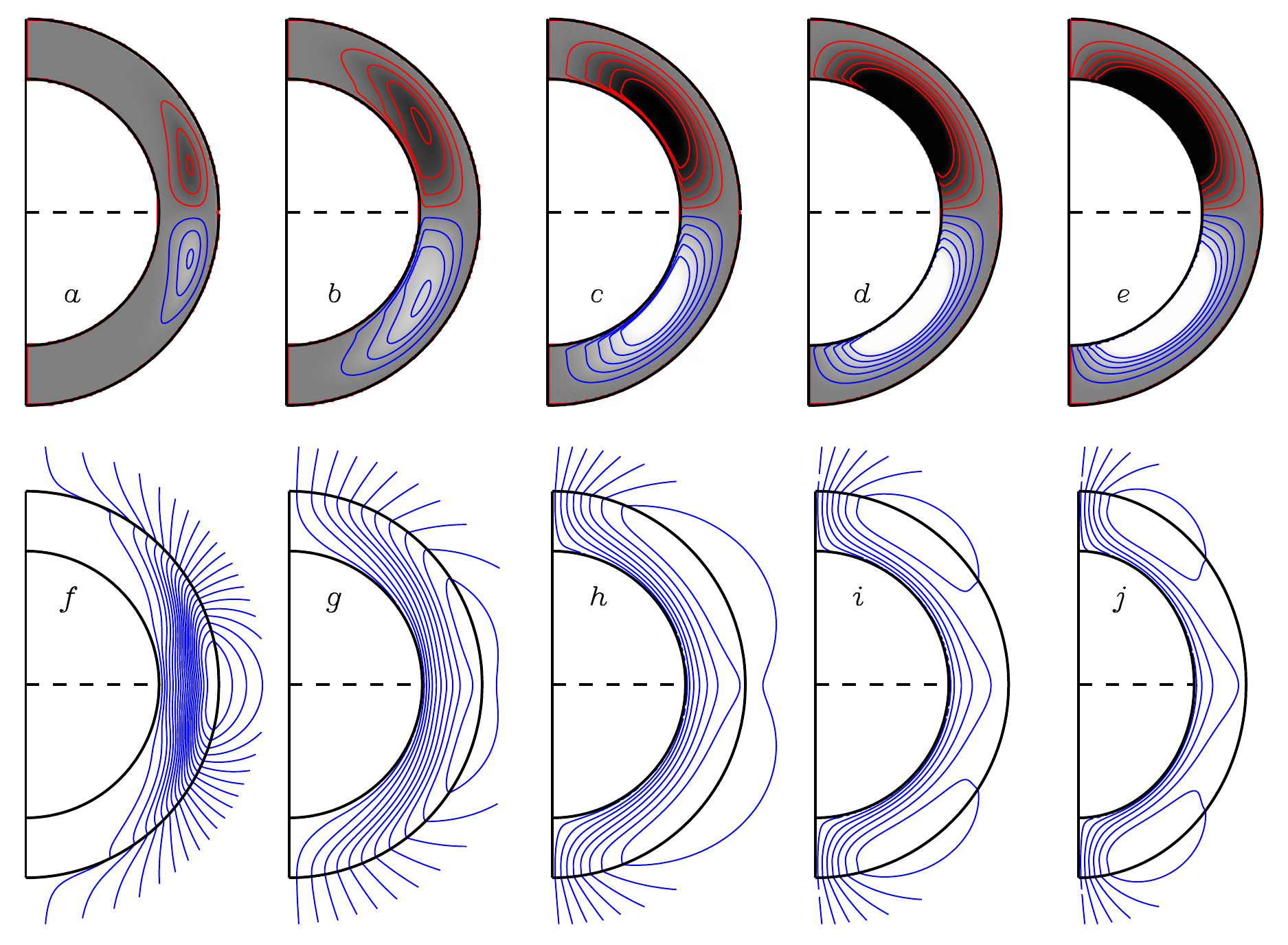}}
\caption{Snapshots of axisymmetric toroidal field lines [(a)–(e)] and axisymmetric poloidal field lines [(f)–(j)] are shown at 5 different times -- (a), (f) = 1.02 yr, (b), (g) = 3.05 yr, (c), (h) = 5.08 yr, (d), (i) = 7.11 yr, and (e), (j) = 9.15 yr. Line contours in frames (a)–(e) show $\hat{B_\phi}$ (azimuthal averaged) with red and blue indicating eastward and westward fields respectively. Filled contour represents strength of the mean toroidal fields (color scale = $\pm$ 1.5 G). The square root of poloidal magnetic potential with potential field extrapolation above the surface (up to r = 1.25R) is shown in frame (f)-(j). Blue color contours denote the clockwise direction of the field. Contour levels corresponding to the poloidal fields strengths of $\pm$ 0.02 G are fixed as maximum and minimum respectively. Taken from \citet{HCM17}.}
\label{fig:field_lines}
\end{figure*}

The `Spotmaker' algorithm (as explained in Section~\ref{sec:buoyancy} is used to place two sunspots pairs symmetrically across equator at two hemispheres at $\pm$ 5$^{\circ}$ latitudes. The magnetic flux in each spot is chosen as 1 $\times$ 10$^{22}$ ${\rm Mx}$  and its radius is taken to be 21.71 ${\rm Mm}$. To make result of our simulation more clearly visible, the radius of each sunspot is chosen somewhat larger than actual sunspot. After placing the sunspots successfully, we allow our code to evolve the magnetic field from these sunspot pairs leading to the build-up of the polar field. The snapshots of radial magnetic field ($B_r$) at different times during evolution of the magnetic field is shown in Figure~\ref{fig:twopairs}. Figure~\ref{fig:field_lines} shows the evolution of toroidal field and poloidal field at different times after the initial placement of the sunspot pairs. As soon as a sunspot pair is placed using `Spotmaker' algorithm, some toroidal field arises below the surface at once because the magnetic loop connecting two sunspots below have a toroidal component. Also, more toroidal field is generated because of the latitudinal differential rotation in the convection zone.

If the two pairs of sunspots are sufficiently close to the equator, then leading polarity sunspots from both hemispheres get canceled by diffusion across equator. The trailing polarity sunspots are preferentially transported to the higher latitudes and get stretched by differential rotation. The meridional circulation takes almost 3 years to bring flux of $B_r$ to produce a positive patch at northern hemisphere and a negative patch in the southern hemisphere as shown in Figure~\ref{fig:twopairs}. The polar magnetic patches form with the polarity of the  trailing sunspots. However, a careful look at Figure~\ref{fig:twopairs} shows some evidence of opposite polarity of what we see in the poles at mid-latitudes even  two sunspots are placed symmetrically at sufficiently low latitudes in both the hemispheres. To understand the physics of what is happening, we need to focus on the poloidal field lines plot in bottom panel of Figure~\ref{fig:field_lines}(f)-(g). After magnetic fluxes from the leading sunspots near equator cancel out, we get initial poloidal field lines spanning both the hemispheres. As it is clear from the early stage of magnetic field evolution (Figure~\ref{fig:field_lines}(f)), we have $B_r$ only at high latitudes. In the later stage, when meridional circulation drags the poloidal field lines towards poles, eventually polar fields in two hemispheres get detached, as a result of which $B_r$ again appears at lower latitudes having the opposite polarity of $B_r$ at high latitudes.

In the 2D SFT model, only the fluxes from the following polarities are advected towards the poles and we eventually get polar patches that are not surrounded by the opposite polarity patches, as found in the 3D models. The outward spreading magnetic field from the polar patches by diffusion is eventually balanced by the inward  advection by the meridional circulation and an asymptotic steady state is reached in the SFT model, with an asymptotic magnetic dipole that does not change with time (see  Figure~6 of \citet{Jiang14}). But for the 3D case, different scenario arises because of the 3D structure of the magnetic field. As $\nabla. {\bf B} = 0$, $\int {\bf B}.d{\bf S}$ integrated over the whole surface has to be zero at any time. This means that during any time interval, equal amounts of positive and negative magnetic fluxes need to disappear below the surface due to subduction process. Hence, The low latitude emergence of the $B_r$ is due to the 3D structure of the magnetic fields. Since full vectorial nature is not considered in the SFT model, low latitudes $B_r$ would never appear in that model. Because breakup of the poloidal field and the appearance of $B_r$ at the low latitudes with opposite polarity, it is possible for the poloidal field to be subducted below the surface in the two hemispheres as the meridional circulation sinks downward in the polar regions. Hence, in contrast to the SFT models where polar field have nothing to cancel them and therefore persist, the polar field disappears after some time in the 3D model.

\begin{figure}[htbp!]
\centering{\includegraphics[width=1.0\columnwidth]{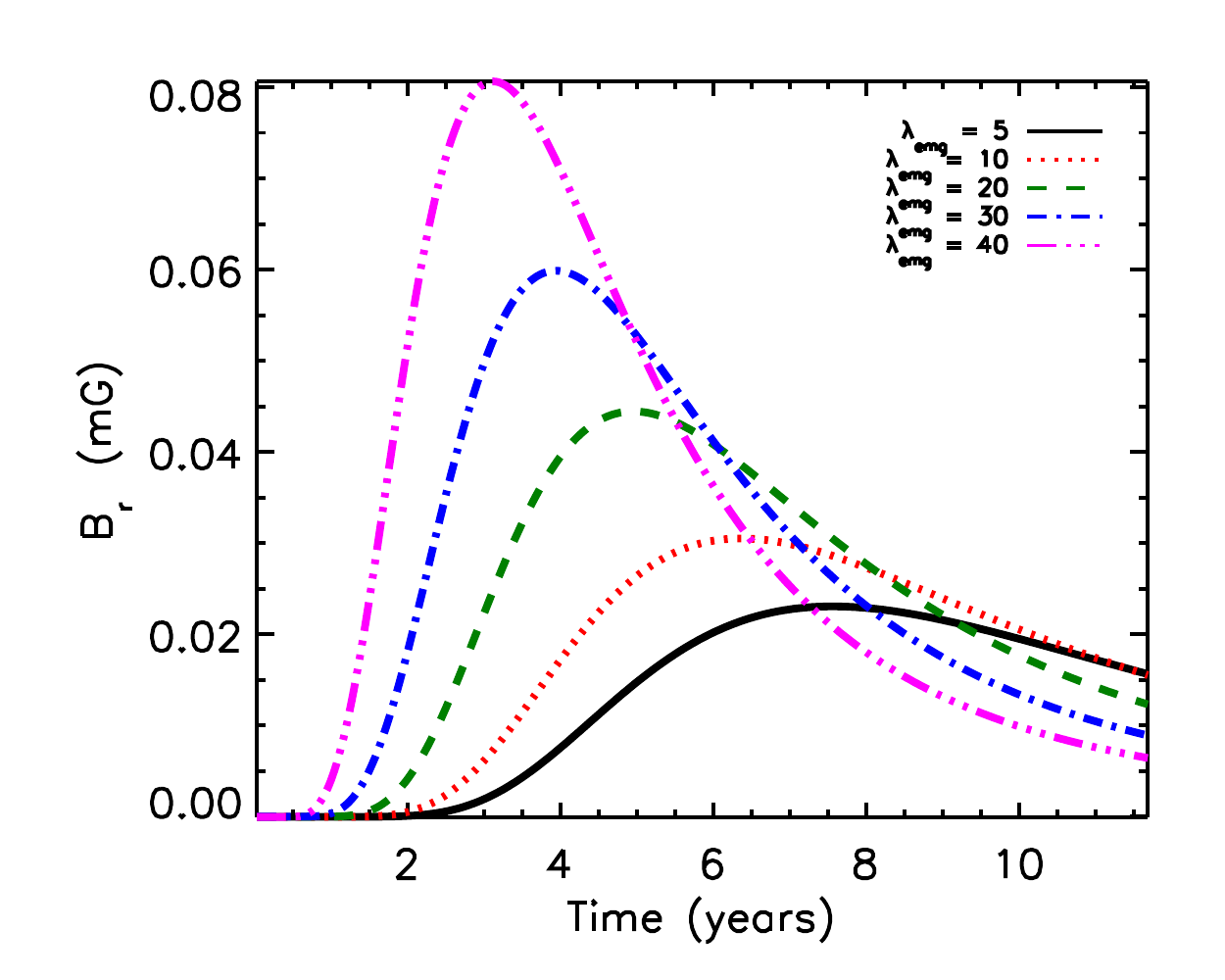}}
\caption{ Time evolution of polar field for different emergence angle ($\lambda_{\rm emg}$) of sunspot pairs in both hemispheres from \citet{HCM17}. Black solid, red dotted, green dashed, blue dash dotted, and magenta long dash dotted lines show the polar field for the sunspot emergence at 5$^{\circ}$, 10$^{\circ}$, 20$^{\circ}$, 30$^{\circ}$, and 40$^{\circ}$ respectively. The unit of magnetic field is in milliGauss and time is given in years.}
\label{fig:pfield}
\end{figure}

\begin{figure*}[htbp!]
\centering{\includegraphics[width=.68\textheight]{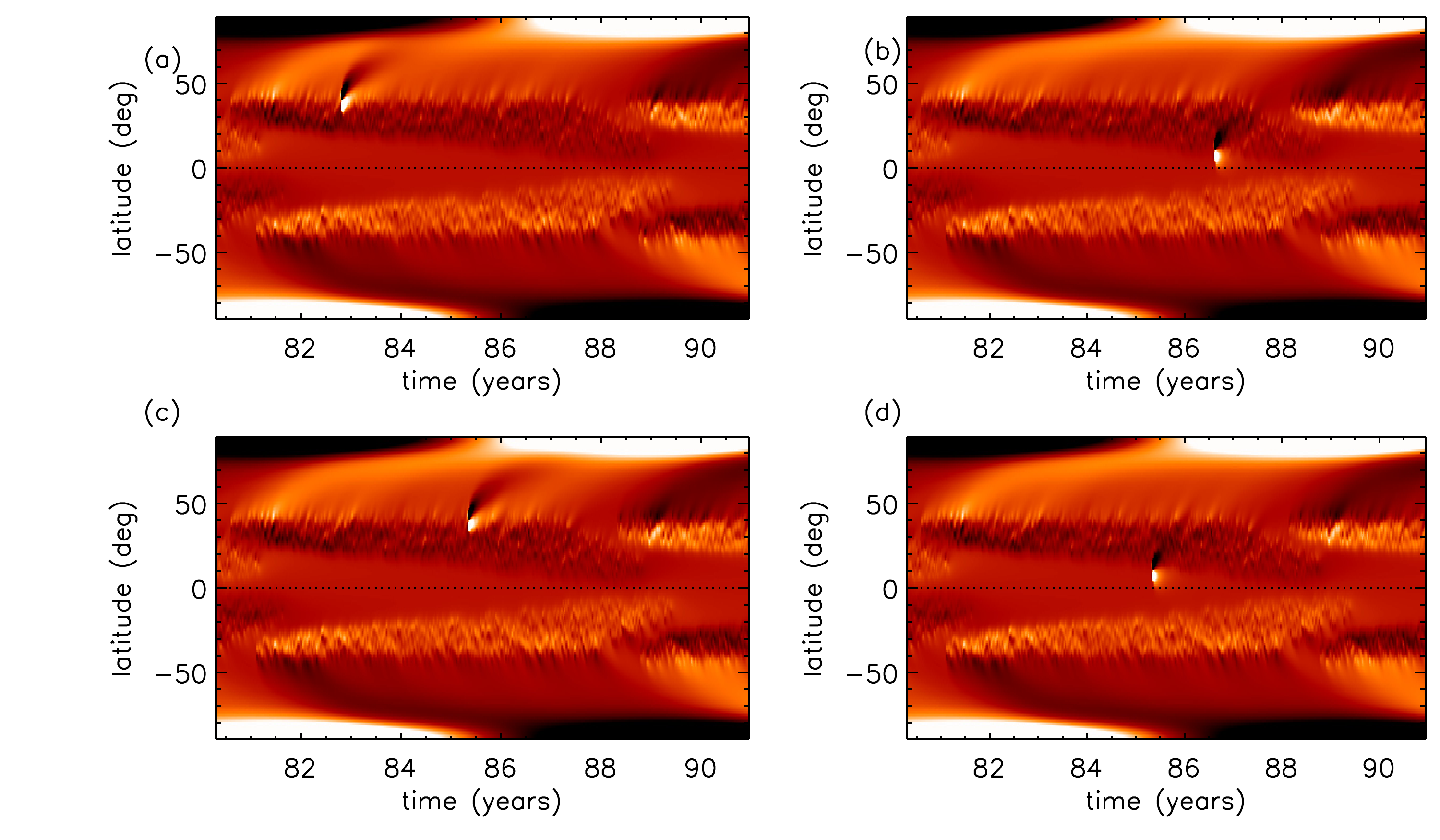}}
\caption{Time-latitude diagram of the radial field on the solar surface with an ``anti-Hale” sunspot pair at different latitudes and different phases of the solar cycle. (a) At early phase of the cycle and at 40$^\circ$ latitude, (b) late phase of the cycle and 10$^{\circ}$ latitude, (c) middle phase and 40$^{\circ}$, and (d) middle phase and 10$^{\circ}$ latitude. Color scale saturates at $\pm$ 15 kG for all four cases. Taken from \citet{HCM17}.}
\label{fig:bfly_ahale}
\end{figure*}

Next, we present results by placing two sunspot pairs symmetrically at different latitudes in the two hemispheres. Evolution of the polar field for sunspot pairs placed at different latitudes is shown in the Figure~\ref{fig:pfield}. The sunspot pairs that are placed at high latitudes advects less path to the poles and lost less flux due to diffusion and as a result, the polar field becomes stronger for high latitudes of emergence. Eventually, polar field disappears in all cases due to subduction by the meridional flows and diffusion. Figure~\ref{fig:pfield} can be directly compared with the left panel of Figure~6 of \citet{Jiang14} where the time evolution of axial dipole moments are shown. This comparison makes the difference between the 3D model and SFT model completely clear. In the SFT model, the cross equator diffusion for the sunspot pairs that are put at sufficiently high latitudes is negligible and fluxes of both polarity are advected to the polar region and eventually the axial dipole moment becomes zero. In case of sunspots pair placed at low latitudes in the SFT model, only the fluxes from the following polarity sunspots reach to the poles and give rise to an asymptotic axial dipole. Situation becomes completely different in the 3D model. However, we see the persistence of polar field for longer time when the inital sunspots are placed at lower latitudes. The sunspot pairs appearing in lower latitudes is somewhat more effective in creating the poloidal field even in the 3D model. This finding is also in agreement with the results of \citet{Dasiespuig10} who found a better correlation between the average tilt of a cycle and the strength of the next cycle if more weight is given to sunspot pairs at low latitudes while computing the average tilt.

Although, the difference between the result of SFT model and 3D model is notable, it may be offset to some extent by including efficeint downward radial pumping. \citet{KC16} have shown that downward radial pumping due to strongly stratified convection near the solar surface can suppress the upward diffusion of toroidal and poloidal fields. Hence, turbulent pumping can help to produce a steady polar field that might persist indefinitely. The exact role of this turbulent pumping needs to be investigated thoroughly.

\subsection{Contribution of anti-Hale sunspot pairs on the polar field}
Since the build up of polar field is much realistically captured in the 3D model compared to the SFT models, we now address another important question that whether the anti-Hale sunspot pairs have a large effect on the polar field in the 3D dynamo model. It is well known that some of the bipolar magnetic regions appear on the solar surface with wrong magnetic polarities not obeying Hale's polarity law. While flux tubes rise through the solar convection zone, it gets affected by the action of turbulence \citep{Longcope02,Weber11}. As a result we see a spread of tilt angles around Joy's law \citep{Hale19}. Due to the spread in tilt angles around Joy's law, it is quite expected that a few outliers would violate Hale's law. A study by \citet{SK12} estimated about 4\% of medium and large sunspots violate Hale's law. Since this is a small percentage of sunspot numbers, it is not surprising that due to statistical fluctuation, these anti-Hale sunspots appear in some particular cycles compared to the other cycles. \citet{Jiang15} suggested on the basis of their SFT calculations that appearance of a few large anti-Hale sunspot pairs at a particular cycle can significantly decrease the strength of polar field at the end of that cycle and this is the reason of weak polar field at the end of cycle 23 but not cycle 21 or 22.

\begin{figure}[htbp!]
\centering{\includegraphics[width=0.95\columnwidth]{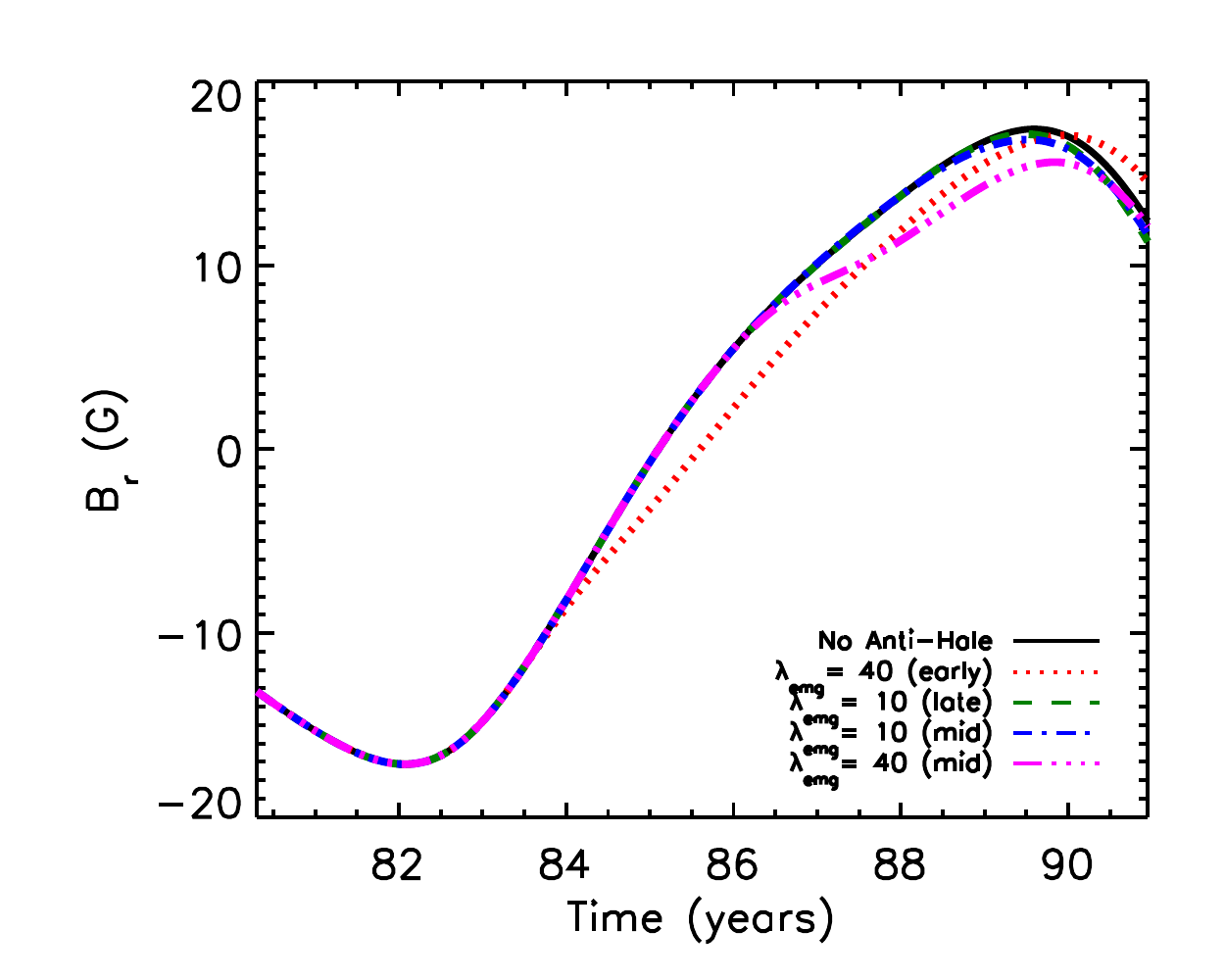}}
\caption{Time evolution of polar field for one complete solar cycle with the ``anti-Hale" sunspot pair at different locations and different times of the cycle taken from \citet{HCM17}. A case of regular cycle with no anti-Hale sunspot pair is plotted in solid black line. Red dotted line shows the poloidal field evolution with an anti-Hale pair at 40$^{\circ}$ latitude at an early phase of the cycle. Green dashed line indicates poloidal field with an anti-Hale pair at 10$^{\circ}$ and late phase. Blue dashed and magenta long dashed lines represent the poloidal field with an anti-Hale pair at the middle of the cycle but at 10$^{\circ}$ and at 40$^{\circ}$ latitudes respectively.}
\label{fig:pfield_ahale}
\end{figure}

A large anti-Hale sunspot pair is placed manually by hand in different phases during a cycle in the 3D and its effect on the build up of polar field is studied in the 3D model. The magnetic flux in the anti-Hale sunspot pairs is chosen 25 times the magnetic flux carried by other regular sunspots to make its effect more clearly visible. The tilt angle for the anti-Hale pair is taken as 30$^{\circ}$. We consider four different cases to understand how appearance of an anti-Hale sunspot pair at different emergence latitudes and different phases of the cycle effects the polar field. As sunspots generally appear at high latitudes in the early phase of the cycle and at low latitudes in the late phase, we consider one case by placing the anti-Hale sunspot pair at the high latitude of 40$^{\circ}$ in the early phase and another case by putting the anti-Hale pair in the late phase at 10$^{\circ}$. In the other two cases, the anti-Hale sunspot pair is placed at 40$^{\circ}$ and 10$^{\circ}$ latitudes (different case studies) in the middle phase of the cycle. Figures~\ref{fig:bfly_ahale} shows the time-latitude plot (``butterfly diagram") of radial magnetic field for these four cases. The time evolution of the polar field in these four cases along with a case without an anti-Hale sunspot pair is shown in Figure~\ref{fig:pfield_ahale}, where the effect of the anti-Hale pair on the polar field is clearly visible.

It is found that an anti-Hale pair near the low-latitude like 10$^{\circ}$ at any phase of the cycle does not have much effect on the polar field as it is clear from Figure~\ref{fig:pfield_ahale}. The opposite fluxes from the two sunspots neutralize each other before they reach the poles, which becomes quite apparent from the figures~\ref{fig:bfly_ahale}(b) and (d). The anti-Hale pair at low-latitudes produce a surge behind them but it does not reach to the pole. If the anti-hale pair appears at the high latitudes, its effect is certainly much more pronounced. The surges in these cases reach to the poles as we see in figures \ref{fig:bfly_ahale}(a) and (c). When anti-Hale pair appears at 40$^{\circ}$ but at early phase of the cycle, the build up of polar field is weakened and delayed, but eventually polar field reaches almost the strength of the polar field that we expect in absence of the anti-Hale sunspot pair (Figure~\ref{fig:pfield_ahale}). However, if that pair appears in the middle phase of a cycle, the polar field can be reduced by about 17\%. Note that this large reduction arises because the anti-Hale sunspot pair is unrealistically large. In conclusion, an anti-Hale sunspot pair could affect the build up of the polar field--especially if they appear at high latitudes and in the middle phase of a cycle but the effect does not appear to be very dramatic.

On the other hand, recent calculation by \citet{Nagy17} using their 2 $\times$ 2D hybrid dynamo model showed that an individual large anti-Hale pair appear as far as 20$^{\circ}$ from the equator can still have a significant effect on the polar field. The strongest effect on the subsequent cycle occurs when a large pair emerges around cycle maxima but at low latitudes. This finding is also in accordance with the \citet{Jiang15} that suggested the weakness of the polar field at the end of cycle 23 was due to the appearance of several anti-Hale sunspot pairs. Since some of the results from 3D model differs with SFT models due to low latitude poloidal flux emergence, we believe the difference occurs in the result of effect of anti-Hale pair on the polar field in our model and other models is a consequence of the same low-latitude poloidal flux emergence. However, This suggestion merits further detailed study in order to arrive a firm conclusion.

\begin{figure*}[htbp!]
\centering{\includegraphics[width=.75\textheight]{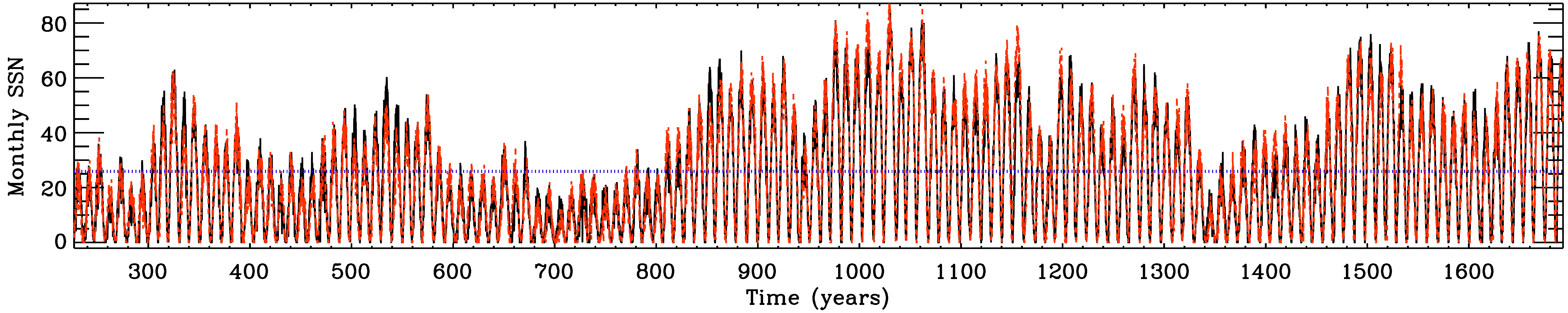}}
\caption{Smoothed (over three months) monthly sunspot number is plotted with time. Black and red lines show north and south sunspot numbers respectively. The dotted line represents the mean of peak sunspot numbers for the last 13 observed solar cycles. This figure is taken form \citet{KM17}.}
\label{fig:irregular}
\end{figure*}

\section{Irregularities of the Solar cycle from 3D models}
The 3D dynamo model is used to study irregularity of solar cycle as well. The plausible causes of irregularity in the solar cycle in the BL dynamo framework include variation in different flux transport mechanisms (convective transport and transport by meridional circulation), differential rotation and randomness in the BL process. Modeling stochastic variation in convective flows is a challenging problem. It needs unified understanding of small and large scale dynamo action and has been studied by a few groups (e.g., \citet{Kitchatinov94,Karak14}). The influence of variation in meridional flow is found to be very important to give rise of cycle variability including grand minima and grand maxima \citep{CD2000, KarakChou11, KarakChou13, HKBC15,HC19}. From helioseismolgy, a weak variation in the differential rotation is known to exist (see \citet{CCC09} and reference therein for detail) but the observed correlation between polar flux at the cycle minimum and the sunspot number of the following cycle suggests that the $\Omega$-effect is largely linear and not a major source of irregularity in the solar cycle \citep{Jiang07, Wang09}.

 Direct observations of polar field for a last few cycles \citep{SCK05}, as well as some polar proxies such as polar faculae, Polar Network Index available for the last 100 years show a clear strong cycle-to-cycle variations in the polar field \citep{Munoz13,Priyal14,HC19}. The amount of polar field generation mainly depends on the tilt angle between Bipolar Magnetic Regions (BMRs), their magnetic fluxes and speed of the meridional circulation. Particularly the scatter of tilt angles around mean, presumably caused by the effect of convective turbulence on the rising flux tubes plays a major role. Recently, \citet{Jiang14} studied that the tilt angle scatter led to a variation of the polar field by about 30\% for cycle 17. Hence, the random scatter in active regions tilt is considered as a possible mechanism to explain the irregularity of the solar cycle \citep{Chou92, CD2000, CCJ07}

 Randoms scatter in the Bipolar magnetic region (BMR) tilt angles has been studied previously within the context of 2D BL dynamo models \citep{Chou92, CD2000,Jiang07,CK09,HKBC15}, SFT models \citep{Jiang_review15} and in a coupled 2 $\times$ 2D BL/SFT model \citep{Lemerle17}. For the first time, \citet{KM17} have considered the random scatter in the tilt angles in the STABLE 3D dynamo model framework. In STABLE model, the standard Joy's law is used for tilt angle $\delta$ = $\delta_0 \cos\theta$, while implementing Spotmaker algorithm to put bipolar sunspots on the surface of the Sun. To include tilt angle scatter around its mean, a random fluctuating component ($\delta_f$) is added around Joy's law as

\begin{equation}
\delta = \delta_0 \cos\theta + \delta_f
\end{equation}

According to observations, Joy's law is a statistical law and there is a considerable scatter around it \citep{Howard91,SK12}. By analysing BMRs data measured during 1976-2008, \citet{Wang15} reported that the fluctuation of the tilt ($\delta_f$) roughly follow a Gaussian distribution as given below:

\begin{equation}
f(\delta_f) = \frac{1}{\sigma_\delta\sqrt{2\pi}}{\rm exp}[-\delta_f^2/2\sigma_\delta^2]
\end{equation}
where $\sigma_\delta$ = 15$^{\circ}$.

Also, \citet{KM17} implemented a tilt angle quenching as a main source of dynamo quenching instead of flux quenching (Equation~\ref{eq:quench1}) as used in previous STABLE papers \citep{MD14,MT16, HCM17,HM18}.
Finally the tilt used to study the irregularity is given by
\begin{equation}
\delta = \frac{\delta_0 \cos\theta + \delta_f}{1 + B(\theta,\phi,t)/B_q^2}
\end{equation}
where $\delta_0$ = 35$^{\circ}$ and others term are as usual. The simulated solar cycle i.e., the sunspot time series (smoothed over three months) is shown in figure~\ref{fig:irregular}. Black and Red represent the sunspot numbers in northern and southern hemispheres. The cycle to cycle variation of the amplitude of the mean polar flux is $\sim$ 35\%. This result is in agreement with the study of  \citet{Jiang14}, who found a 30\% variation in polar field after introducing tilt angle scatter.

The strength of the magnetic field and number of bipolar sunspots per cycle have increased in compare to the case without tilt angle scattering. The variation in the peak SSN in this simulation is 41\%, while the observed variation during the period of 1749-2017 in sunspot number data is 32\%. Also, the hemispheric asymmetry is observed in the simulated time series of the sunspot numbers. By zooming in the figure~\ref{fig:irregular} carefully, one can clearly find a temporal lag and excess of sunspots numbers between two hemispheres. However, like real Sun, STABLE model always tends to correct any hemispheric or temporal asymmetry produced in a cycle and no extended asymmetry is seen. The similar results were also reported using 2D models by \citet{CC06}.

In this 3D model, the cycle period has considerable variation around its mean of 10.5 years. In the flux transport dynamo, cycle period is largely determined by the speed of the meridional flow. But while introducing scatter in tilt angle, the speed of meridional flow is kept constant. Hence, the variation in cycle period occured due to the fluctuation and nonlinearities in the BMR emergence. Basically, when polar field of a cycle becomes stronger due to the tilt fluctuations, spots in the next cycle take a longer time to reverse the previous cycle flux which makes a longer cycle. On the other hand, a stronger polar field makes toroidal field stronger which leads to more frequent BMR emergence. This effect acts to reverse polar field quickly and makes the cycle period shorter, although it is inhibited by the tilt angle quenching. Therefore the tilt angle quenching makes a major role in deciding which effect is prominent and how period will be varied. Hence the introduction of observed tilt angle scatter in the Sun may be sufficient to account for the irregularities observed in the solar cycle.

In 2D BL dynamo models, some of features of the solar cycle (e.g., Waldmeier effect and correlation between decay rate and amplitude of the next cycle) is not reproduced by introducing randomness in the BL process. So, a variation in the meridional circulation was necessary to explain these features. In the 3D dynamo model, a small correlation arises between the cycle amplitude and the period of the previous cycle (r = -0.24) by only including tilt angle scatter, while the observed correlation is r = -0.67.  Hence it needs a separate study whether other irregular properties of the solar cycle can be recovered by introducing tilt angle scatter around joy's law or we need to introduce fluctuation in the meridional circulation as well.

The figure~\ref{fig:irregular} which includes a tilt angle scatter following a gaussian distribution with $\sigma_\delta$ = 15$^{\circ}$ produces very weak and strong cycles and a few Dalton-like extended periods of weak activity (e.g., around 700 years). But it does not produce any Maunder-like grand minimum or grand maximum. Keeping the possibility in mind that the weaker BMRs might have bigger scatter in their tilt and study of tilt angles does show a cycle-to-cycle variations in the tilt angle \citep{SK12, Lemerle15, Jiang14}, the scatter distribution around mean tilt is doubled. Now $\sigma_\delta$ = 30$^{\circ}$ is considered instead of 15$^{\circ}$. Interestingly, this large fluctuation in the tilt angle does not effect the dynamo operation and dynamo never dies. A cyclic behaviour is always maintained. A several episodes of weak magnetic activity i.e., maunder-like grand minima arises from this simulation. This simulation produces occasional periods of stronger magnetic activity or grand maxima as well (see \citet{KM17} for more details).

\begin{figure*}[htbp!]
\centering{\includegraphics[width=.75\textheight]{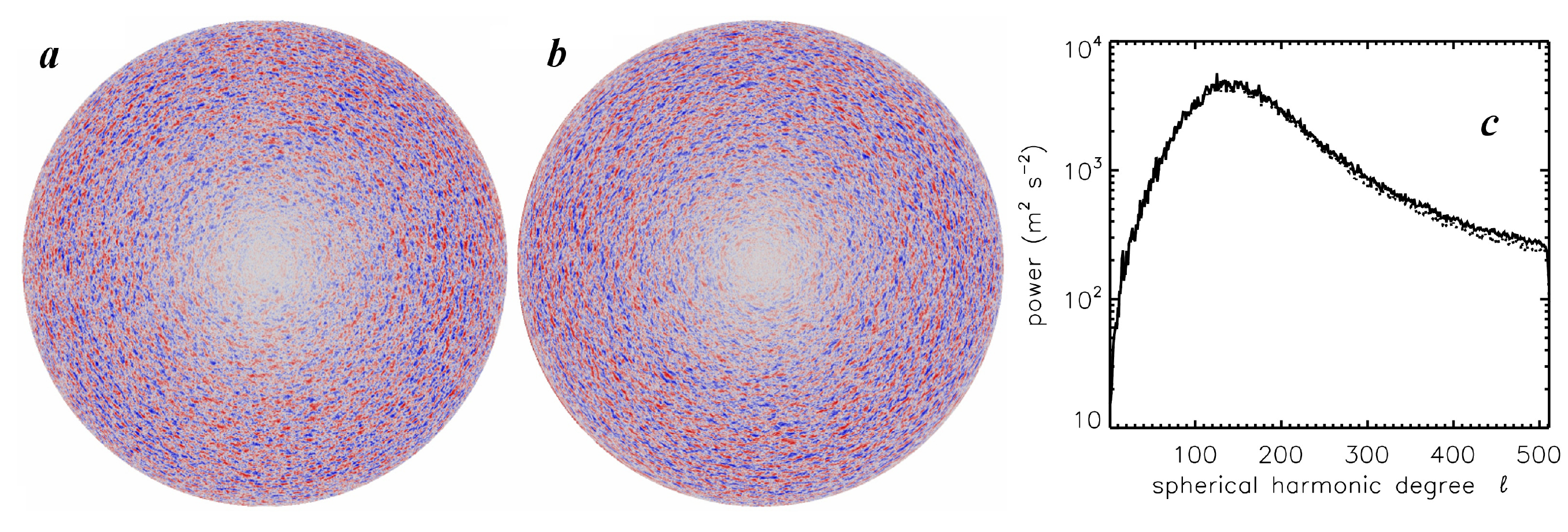}}
\caption{(a) Line-of-sight (Doppler) velocity on the solar photosphere measured using the SOHO/MDI instrument on 1996 June 4. (b) Simulated line-of-sight velocity obtained from the empirical model developed by Hathaway et al. (2000) and Hathaway (2012a, 2012b). Hathaway's models are designed to reproduce the observed horizontal Doppler velocity power spectrum with randomized phases. (c) Horizontal velocity spectra for the convective flow field used here (dotted line) and for Hathaway’s simulated flow field (solid line) at r= R$_\odot$. Small discrepancies at large $\ell$ arise because we only extract the divergent component. Taken from \citet{HM18}.}
\label{fig:cflow_obs}
\end{figure*}

\begin{figure*}[htbp!]
\centering{\includegraphics[width=.75\textheight]{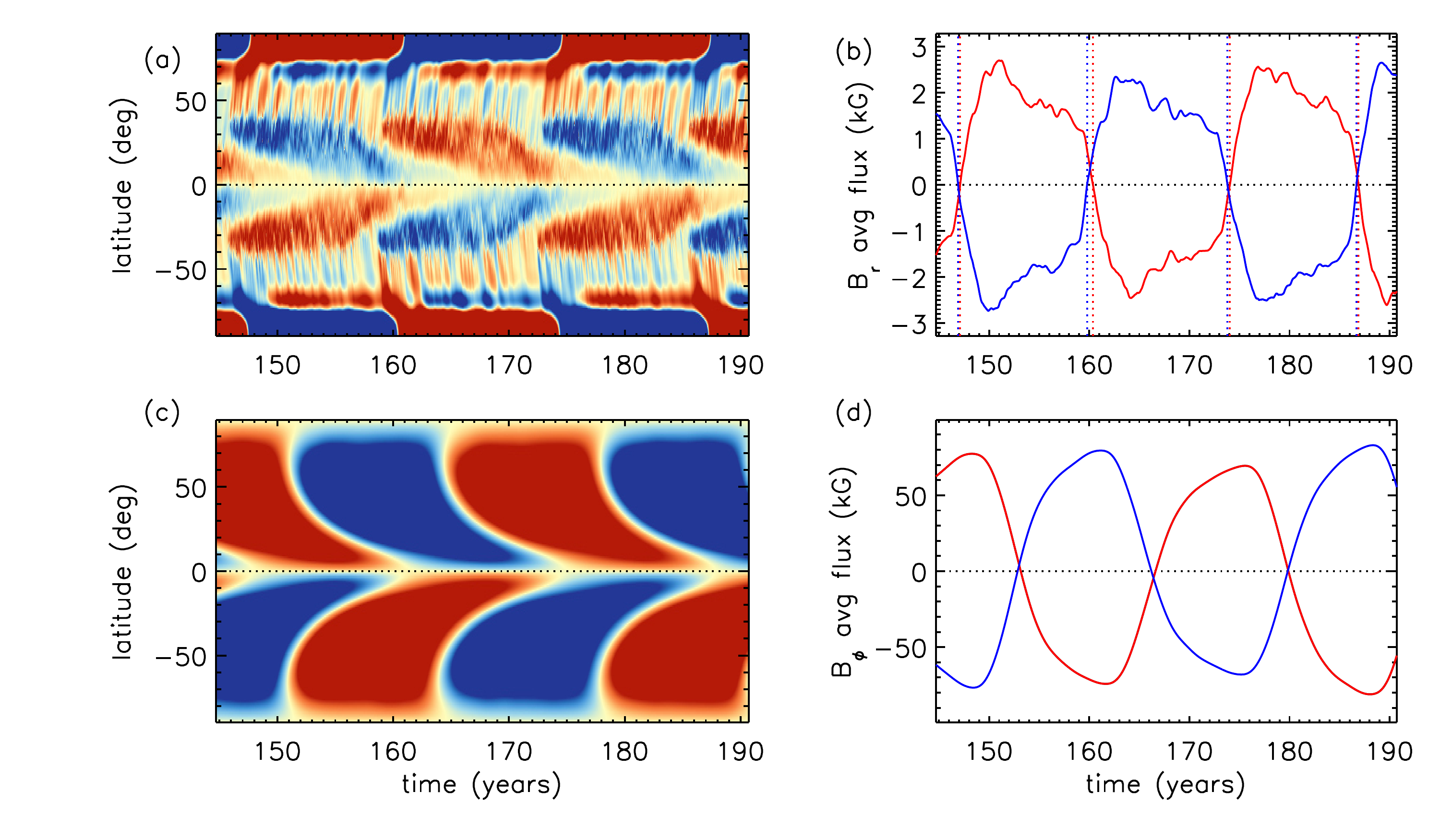}}
\caption{Same as Figure~\ref{fig:bfly} but for the case that includes surface convective flows. The color scale saturates at (a) $\pm$ 500 G and (c) $\pm$ 100 kG. From \citet{HM18}.}
\label{fig:bfly_cflow}
\end{figure*}

\begin{figure*}[htbp!]
\centering{\includegraphics[width=.7\textheight]{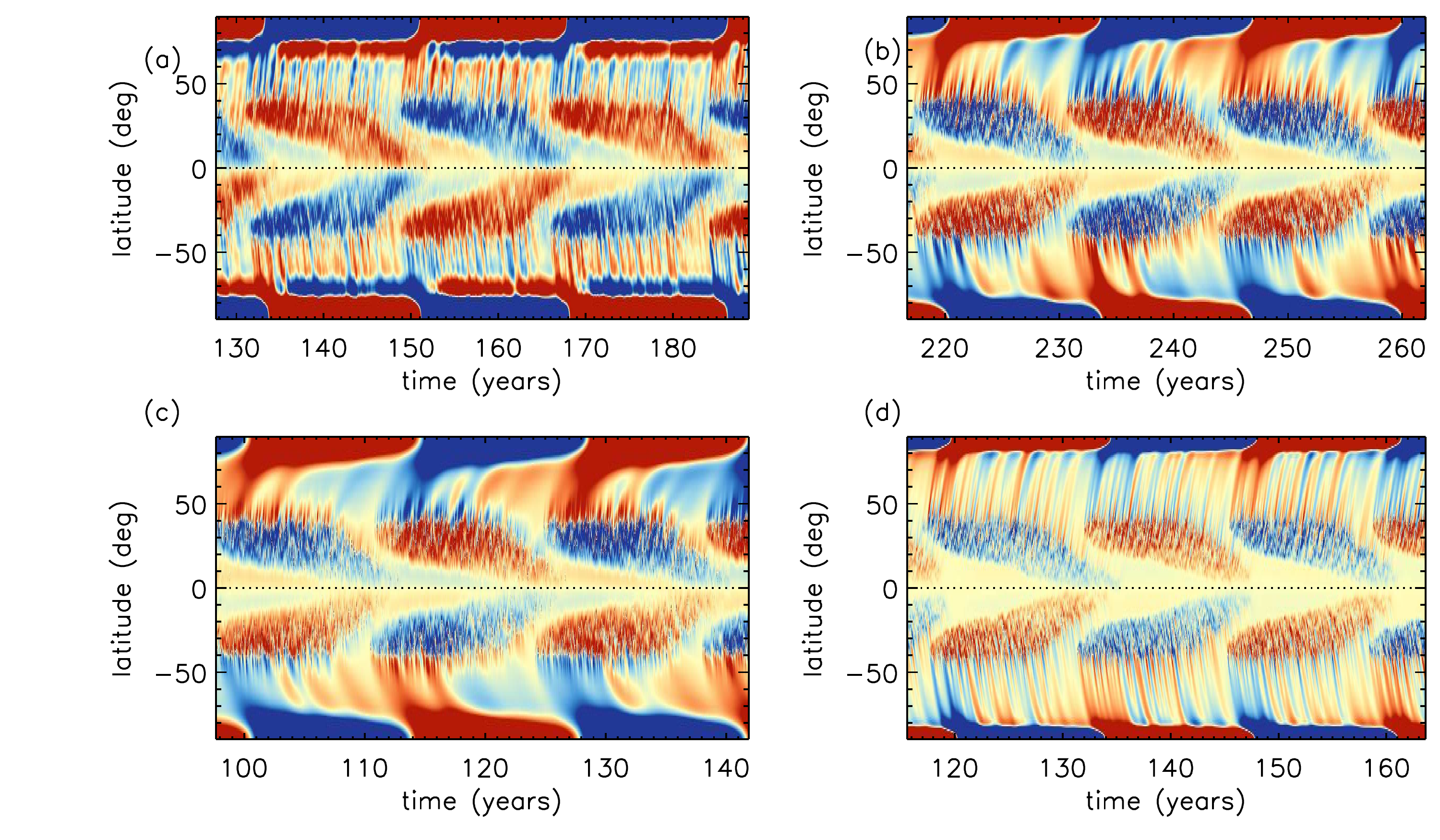}}
\caption{Butterfly diagram of azimuthal averaged $B_r$ at r = R$_\odot$ for (a) Convective case that includes surface convective flows (Br scale = $\pm$ 500 G), (b) Case with $\eta_{top}$= 3 $\times$ 10$^{12}$ cm$^2$ s$^{−1}$ (Br scale = $\pm$ 200 G), (c) Case with
 $\eta_{top}$= 10$^{13}$ cm$^2$ s$^{−1}$ (Br scale = $\pm$ 30 G), and (d) Case with
 $\eta_{top}$= 8 $\times$ 10$^{11}$ cm$^2$ s$^{-1}$ (Br scale = $\pm$ 800 G). From \citet{HM18}.}
\label{fig:bfly_all}
\end{figure*}

\section{Incorporation of the surface convection}
The advantage of extra dimension in the 3D dynamo model allows to explore implementation of overwhelming observational data available for the Sun. In STABLE model, \citet{HM18} incorporate three-dimensional convective flow fields for the first time in a BL dynamo framework to study their effect on the solar cycle. The dispersal and migration of surface fields is modeled as an effective turbulent diffusion in most of the BL models, but they have incorporated the realistic convective flow to capture the BL process more realistically by exploiting 3D capabilities of STABLE model.

The observed line of sight velocity on the photosphere measured with the SOHO/MDI Dopplergram instrument is included in the model. 2D maps of line-of-sight velocity component measured using MDI Dopplergram is shown in figure~\ref{fig:cflow_obs}(a). These velocities in the photosphere are predominantly horizontal and includes all information of Differential rotation, Meridional circulation, other unwanted signals such as convective blueshift, spacecraft motion and instrumental artifacts. In order to reconstruct the complete 2D convective spectrum o the surface of the Sun, one needs to model the data to get rid of all unwanted information. \citet{Hathaway12a, Hathaway12} has devised such a model that subtracted all unwanted above mentioned signals and computed the convective power spectrum. After that a horizontal velocity field is calculated based on that observed spectrum using a series of spherical harmonic with randomized phases. A sample of surface Dopplergram from the simulated horizontal velocity field is shown in figure~\ref{fig:cflow_obs}(b). The power spectrum of the horizontal velocity is also shown in figure~~\ref{fig:cflow_obs}(c). The peak around $l$ = $130$ shows supergranulation.

In order to incorporate the empirical surface velocity field ${\bf V}_s(\theta, \phi, t)$ (figure~\ref{fig:cflow_obs}(b)) into STABLE, a radial structure is specified and the surface field is extrapolated downward into the convection zone no deeper than $r$=$0.9R_\odot$. As convection is very vigorous near the surface layers than the deeper convection zone, the convective velocity fields kept confined near the surface layers. Also, these velocities are non-axisymmetric and they affect the transport and amplification of magnetic field, which can influence the time evolution of the mean fields.

Note that global convective simulations are fully capable of simulating global-scale convective motions but no global simulation model can accurately capture smaller-scale convective motions near the surface of the Sun such as granulation and supergranulation. This is because that would require both extremely high resolution and and some very important physical process such as non-LTE radiative transfer, ionization and full compressibility. However, these small scale convective motions are most important to the breakup and dispersal of BMRs and relevant to the BL process. For this reason, we incorporate convection in a manner that is consistent with the pragmatic approach of the kinematic dynamo modeling.

Soon after implementation of convective flows in STABLE simulation along with other regular ingredients of the model, it is realized that convective flows do operate as a small-scale dynamo disrupting the operation of large scale magnetic cycle. Several approaches are considered to suppress the small-scale dynamo action but only effective way is to make the convective flows acting only on the radial component of the magnetic field. This also helps to realistically capture the horizontal transport of the vertical magnetic field on the surface layers. Also, note that a background diffusivity 5 $\times$
10$^{11}$ cm$^2$s$^{-1}$ is used to maintain dipolar parity. This background diffusivity helps to suppress small scale dynamo as well.

The results with explicit convective transport as the effective transport mechanism of large-scale magnetic field on the surface is shown in figure~\ref{fig:bfly_cflow}. The time evolution of the radial field on the surface (figure~\ref{fig:bfly_cflow}(a)) is comparable with the observed radial field evolution but the presence of mixed polarity near the pole has no counterpart in the observation. The existence of the mixed polarity can be attributed to the tendency for the convective
motions to disperse and migrate BMR field without dissipating them. As a consequence both polarities are migrated towards pole and concentrated into strong, alternating bands.

The polar flux plot in figure~\ref{fig:bfly_cflow}(b) shows somewhat more variability and a sharper decay near the end of each cycle. This variability occurs due to the existence of mixed polarity fields that cross into the polar regions before they cancel each other. The slower decay phase implies a longer interval of polar flux generation by poleward migrating magnetic streams, which persists almost the entire cycle.

The long sustained supply of poloidal flux to the poles throughout most of the cycle has consequences in toroidal field generation. The polar flux behaves as the seed for the next cycle and when it transported to the base of the convection zone by meridional circulation, it promotes sustained toroidal flux generation by $\Omega$-effect, more precisely at the mid-latitudes where the latitudinal shear is strongest. This scenario is evident from figure~\ref{fig:bfly_cflow}(c). Throughout most of the cycle, strong toroidal field persists near $\pm$ 50$^{\circ}$ -60$^\circ$. This existence of strong toroidal flux in the mid-latitude also accounts for the distortion of the integrated toroidal flux curve shown in figure~\ref{fig:bfly_cflow}(d).

\citet{HM18} also address the question whether the convective transport accurately parameterized by a turbulent diffusion or is it fundamentally non-diffusive? In case it is possible to represent convective transport using a turbulent diffusion then what value of surface diffusivity is optimal? In order to answer those questions, a few simulations are performed by changing the surface diffusivity value ranging from 1 $\times$ 10$^{13}$ cm$^2$s$^{-1}$ to 8 $\times$ 10$^{11}$
cm$^2$s$^{-1}$ .

The surface butterfly diagrams for those few cases are shown in figure~\ref{fig:bfly_all}. Qualitatively, the convective case bears the greatest resemblance to the case with surface diffusivity of 3 $\times$ 10$^{11}$ cm$^2$s$^{-1}$. This conclusion is based on the width of the poleward migrating streams, structure of polar field concentration and the relative strength of the polar and low-latitude fields, as well as the location of the active latitudes.

Some more quantitative analysis such as calculation of poleward migration speed, dynamo efficiency shows that a turbulent diffusion coefficient of 3 $\times$ 10$^{11}$ cm$^2$s$^{-1}$ adequately captures the surface flux transport as in the case with explicit convective transport but it does not adequately capture the dissipation of magnetic energy \citep{HM18}. Approximating convective transport with a turbulent diffusion will have an adverse effect on the dynamo efficiency, which in turn will produce artificially weak mean fields and shorter cycles. However, it is found that replacing turbulent diffusion with convective transport does not improve the fidelity to the butterfly diagrams, giving rise to the mixed polarity fields in the polar regions.

\section{Conclusions}
Although many alternative solar dynamo models have been proposed, Babcock-Leighton dynamo model has remained the most promising leading framework to explain various properties of the solar cycle  so far, because it is firmly based on solar observations and provides a robust mechanism to produce cyclic dynamo activity. However, the BL model is kinematic and solves only axisymmetric mean field dynamo equations. The velocity fields are not calculated from the first principle in these models rather they are provided from observations. The basic MHD equations are solved in global MHD models which have progressed a lot recently \citep{Charbonneau14} to reproduce the cyclic activity but they are far away from actual solar parameters. Also, BL model involves many physical process (e.g., BL process, magnetic buoyancy) that rather have been parameterized very simplistic ways \citep{DC99, CNC04, Karak14}. The 3D kinematic BL dynamo models emerge out as a next generation dynamo model that hold the promise to model all physical process more realistically than previous 2D BL models and have capability to include non-kinematic effects such as back reaction due to the Lorentz force on the flows.

Development of such kind of 3D models have been started very recently \citep{YM13, MD14,MT16, HCM17, KM17, Lemerle15, Lemerle17, Kumar19}. The rise of toroidal flux tube due to magnetic buoyancy and subsequent creation of sunspots, which is one of the back bone of BL models, is captured more realistically by  including existing knowledge from 3D MHD flux tube emergence simulations as well as observations. Some authors \citep{YM13, Kumar19} have modeled the buoyant rise of flux tube by applying a radial outward velocity and a vortical velocity simultaneously to a localized part of the a toroidal flux tube near the bottom of the convection zone assuming that parent flux tubes for sunspots remained connected to its root. Whereas others (e.g., STABLE model \citet{MD14, MT16, HCM17}) have assumed that the sunspots get quickly disconnected from the parent flux tube and place the sunspots based on the information of toroidal flux near the base of the convection zone. In this case, a subsurface structure of sunspots is also specified up to radius 0.9 R$_\odot$ from the surface. The latter scenerio is prefered as it is argued by \citet{Longcope02, Rempel05} that sunspots get quickly disconnected from it parent flux tubes. Recently \citet{whitbread19} also argued based on the surface flux evolution that BMRs need to be disconnected from the base of convection zone more rapidly to get better evolution of the surface fields.

The 3D models also study how the solar polar field builds up from the decay of one tilted bipolar sunspot pair and two symmetrically situated sunspot pairs in two hemispheres. It is found that the polar field arises from such sunspots pairs eventually disappears due to the emergence of poloidal flux at low latitudes and subsequent subduction by the meridional flow. This processes are not included in 2D SFT models and the polar field in that models can only be neutralized by diffusion with a field of opposite polarity. So we conclude that one has to be cautious in interpreting the results of 2D SFT models pertaining to polar fields as SFT models do not capture the dynamics of the polar fields more realistically.

How a few large sunspot pairs violating Hale's law could effect the strength of the polar field is also studied in the 3D models. It is found that an anti-Hale pairs do effect solar polar field--especially if they appear at higher latitudes and during mid-phase of the cycle but the effect is not very dramatic.

Irregularities of the solar cycle are also studied in great details using 3D models. In contrast to the previous 2D BL dynamo models where randomness arises in the BL mechanism due to the scatter of tilt angle across Joy's law is included using a stochastic random parameterization as a cause of irregularity, 3D models actually incorporates the scatter of tilt angle physically in the model. Motivated from observational finding, a scatter of tilt around Joy's law is modeled using a gaussian distribution. The cycle irregularities are reproduced quite nicely including grand minima and grand maxima by varying the parameter of the scatter distribution \citep{KM17}.

3D capability of these models is exploited to incorporate high resolution observed data into the dynamo simulations directly to study their effect on solar cycle. In the BL dynamo models, the convective flows on the surface of the Sun plays a key role in the migration and dispersion of the sunspot fields. Usually, these whole procedure are modeled using an effective turbulent diffusion. This might be because of unavailability of enough resolution and framework to incorporate convective flows in the simulation. With the newly developed 3D model, realistic convective flows based on the solar observation is incorporated in order to improve the fidelity of convective transport. This new study shows that approximating  convective transport using a turbulent diffusion underestimates dynamo efficiency producing weaker mean fields and shorter cycle.

3D kinematic dynamo models are capable of producing many attractive features of the solar cycle that were previously not possible by 2D BL dynamo models or 2D SFT models individually. These models incorporates all attractive aspect of both the models, while being free form the limitation of both. The time evolution of the simulated surface magnetic field can be used to study their effect on the topology of corona and on properties of the solar wind. These models also help us to directly assimilate observed surface data and study their interaction with solar interior. The direct incorporation of surface data will help us to predict correctly the polar field and hence in predicting the next cycle amplitude. Presently, a snap-shot of observed convective flows is only incorporated to study the effect of them on the surface evolution of the field and effect on the dynamo cycle. In future, It will be a really important step to assimilate the time-evolving convective flows fields as well as observe surface magnetogram to constrain the polar field at end of the cycle. Presently, a small scale dynamo action after assimilating convective flow fields is encountered, which disrupts large-scale dynamo activity. This small-scale dynamo action can be handle properly if we include Lorentz force feedback to saturate them. Including Lorentz force feedback in the 3D dynamo model would be next step to make these models more suitable to study solar and stellar magnetic cycles.

\section*{Acknowledgments}
The author would like to thank Prof. Arnab Rai Choudhuri for thoroughly going through the manuscript and giving many useful suggestions that help a lot to improve the review. The author also acknowledges the funding from the European Research Council (ERC) under the European Union’s Horizon 2020 research and innovation programme (grant agreement No 817540, ASTROFLOW).

\vspace{-1em}

\bibliography{myref}

\begin{thebibliography}{77}
\providecommand{\natexlab}[1]{#1}
\providecommand{\url}[1]{\texttt{#1}}
\expandafter\ifx\csname urlstyle\endcsname\relax
  \providecommand{\doi}[1]{doi: #1}\else
  \providecommand{\doi}{doi: \begingroup \urlstyle{rm}\Url}\fi

\bibitem[{Parker}(1955)]{Parker55a}
E.~N. {Parker}.
\newblock {Hydromagnetic Dynamo Models.}
\newblock \emph{\apj}, 122:\penalty0 293--314, September 1955.
\newblock \doi{10.1086/146087}.

\bibitem[{Karak} et~al.(2014{\natexlab{a}}){Karak}, {Kitchatinov}, and
  {Choudhuri}]{KKC14}
B.~B. {Karak}, L.~L. {Kitchatinov}, and A.~R. {Choudhuri}.
\newblock {A Dynamo Model of Magnetic Activity in Solar-like Stars with
  Different Rotational Velocities}.
\newblock \emph{\apj}, 791:\penalty0 59, August 2014{\natexlab{a}}.
\newblock \doi{10.1088/0004-637X/791/1/59}.

\bibitem[{Hazra} et~al.(2019){Hazra}, {Jiang}, {Karak}, and
  {Kitchatinov}]{HJKK19}
Gopal {Hazra}, Jie {Jiang}, Bidya~Binay {Karak}, and Leonid {Kitchatinov}.
\newblock {Exploring the Cycle Period and Parity of Stellar Magnetic Activity
  with Dynamo Modeling}.
\newblock \emph{\apj}, 884\penalty0 (1):\penalty0 35, October 2019.
\newblock \doi{10.3847/1538-4357/ab4128}.

\bibitem[{Hazra} et~al.(2020){Hazra}, {Vidotto}, and {D'Angelo}]{HVD20}
Gopal {Hazra}, Aline~A. {Vidotto}, and Carolina~Villarreal {D'Angelo}.
\newblock {Influence of the Sun-like magnetic cycle on exoplanetary atmospheric
  escape}.
\newblock \emph{\mnras}, 496\penalty0 (3):\penalty0 4017--4031, June 2020.
\newblock \doi{10.1093/mnras/staa1815}.

\bibitem[{Hale}(1909)]{Hale1909}
G.~E. {Hale}.
\newblock {Note on the Magnetic Field in Sun-Spots}.
\newblock \emph{\pasp}, 21:\penalty0 205, October 1909.
\newblock \doi{10.1086/121926}.

\bibitem[{Choudhuri}(1998)]{Chou98}
A.~R. {Choudhuri}.
\newblock \emph{{The physics of fluids and plasmas : an introduction for
  astrophysicists (Cambridge: Cambridge University Press)}}.
\newblock November 1998.

\bibitem[{Choudhuri} et~al.(1995){Choudhuri}, {Sch\"ussler}, and
  {Dikpati}]{CSD95}
A.~R. {Choudhuri}, M.~{Sch\"ussler}, and M.~{Dikpati}.
\newblock {The solar dynamo with meridional circulation.}
\newblock \emph{\aap}, 303:\penalty0 L29--L32, November 1995.

\bibitem[{Durney}(1995)]{Durney95}
B.~R. {Durney}.
\newblock {On a Babcock-Leighton dynamo model with a deep-seated generating
  layer for the toroidal magnetic field}.
\newblock \emph{\solphys}, 160:\penalty0 213--235, September 1995.
\newblock \doi{10.1007/BF00732805}.

\bibitem[{Durney}(1997)]{Durney97}
B.~R. {Durney}.
\newblock {On a Babcock-Leighton Solar Dynamo Model with a Deep-seated
  Generating Layer for the Toroidal Magnetic Field. IV.}
\newblock \emph{\apj}, 486:\penalty0 1065--1077, September 1997.

\bibitem[{Dikpati} and {Charbonneau}(1999)]{DC99}
M.~{Dikpati} and P.~{Charbonneau}.
\newblock {A Babcock-Leighton Flux Transport Dynamo with Solar-like
  Differential Rotation}.
\newblock \emph{\apj}, 518:\penalty0 508--520, June 1999.
\newblock \doi{10.1086/307269}.

\bibitem[{Chatterjee} et~al.(2004){Chatterjee}, {Nandy}, and
  {Choudhuri}]{CNC04}
P.~{Chatterjee}, D.~{Nandy}, and A.~R. {Choudhuri}.
\newblock {Full-sphere simulations of a circulation-dominated solar dynamo:
  Exploring the parity issue}.
\newblock \emph{\aap}, 427:\penalty0 1019--1030, December 2004.
\newblock \doi{10.1051/0004-6361:20041199}.

\bibitem[{Hazra} et~al.(2014){Hazra}, {Karak}, and {Choudhuri}]{HKC14}
G.~{Hazra}, B.~B. {Karak}, and A.~R. {Choudhuri}.
\newblock {Is a Deep One-cell Meridional Circulation Essential for the Flux
  Transport Solar Dynamo?}
\newblock \emph{\apj}, 782:\penalty0 93, February 2014.
\newblock \doi{10.1088/0004-637X/782/2/93}.

\bibitem[{Karak} et~al.(2014{\natexlab{b}}){Karak}, {Rheinhardt},
  {Brandenburg}, {K{\"a}pyl{\"a}}, and {K{\"a}pyl{\"a}}]{Karak14}
B.~B. {Karak}, M.~{Rheinhardt}, A.~{Brandenburg}, P.~J. {K{\"a}pyl{\"a}}, and
  M.~J. {K{\"a}pyl{\"a}}.
\newblock {Quenching and Anisotropy of Hydromagnetic Turbulent Transport}.
\newblock \emph{\apj}, 795:\penalty0 16, November 2014{\natexlab{b}}.
\newblock \doi{10.1088/0004-637X/795/1/16}.

\bibitem[{D'Silva} and {Choudhuri}(1993)]{DSilva93}
S.~{D'Silva} and A.~R. {Choudhuri}.
\newblock {A theoretical model for tilts of bipolar magnetic regions}.
\newblock \emph{\aap}, 272:\penalty0 621--633, May 1993.

\bibitem[{Hale} et~al.(1919){Hale}, {Ellerman}, {Nicholson}, and {Joy}]{Hale19}
G.~E. {Hale}, F.~{Ellerman}, S.~B. {Nicholson}, and A.~H. {Joy}.
\newblock {The Magnetic Polarity of Sun-Spots}.
\newblock \emph{\apj}, 49:\penalty0 153, April 1919.
\newblock \doi{10.1086/142452}.

\bibitem[{Babcock}(1961)]{Bab61}
H.~W. {Babcock}.
\newblock {The Topology of the Sun's Magnetic Field and the 22-YEAR Cycle.}
\newblock \emph{\apj}, 133:\penalty0 572--587, March 1961.
\newblock \doi{10.1086/147060}.

\bibitem[{Leighton}(1969)]{Leighton69}
R.~B. {Leighton}.
\newblock {A Magneto-Kinematic Model of the Solar Cycle}.
\newblock \emph{\apj}, 156:\penalty0 1--26, April 1969.
\newblock \doi{10.1086/149943}.

\bibitem[{Kitchatinov} and {Olemskoy}(2011)]{KO11a}
L.~L. {Kitchatinov} and S.~V. {Olemskoy}.
\newblock {Does the Babcock-Leighton mechanism operate on the Sun?}
\newblock \emph{Astron. Lett.}, 37:\penalty0 656--658, September 2011.
\newblock \doi{10.1134/S0320010811080031}.

\bibitem[{Dasi-Espuig} et~al.(2010){Dasi-Espuig}, {Solanki}, {Krivova},
  {Cameron}, and {Pe{\~n}uela}]{Dasiespuig10}
M.~{Dasi-Espuig}, S.~K. {Solanki}, N.~A. {Krivova}, R.~{Cameron}, and
  T.~{Pe{\~n}uela}.
\newblock {Sunspot group tilt angles and the strength of the solar cycle}.
\newblock \emph{\aap}, 518:\penalty0 A7, July 2010.
\newblock \doi{10.1051/0004-6361/201014301}.

\bibitem[{Wang} et~al.(1989{\natexlab{a}}){Wang}, {Nash}, and {Sheeley}]{WNS89}
Y.-M. {Wang}, A.~G. {Nash}, and N.~R. {Sheeley}, Jr.
\newblock {Magnetic flux transport on the sun}.
\newblock \emph{Science}, 245:\penalty0 712--718, August 1989{\natexlab{a}}.
\newblock \doi{10.1126/science.245.4919.712}.

\bibitem[{Wang} et~al.(1989{\natexlab{b}}){Wang}, {Nash}, and
  {Sheeley}]{WNS89a}
Y.~M. {Wang}, A.~G. {Nash}, and Jr. {Sheeley}, N.~R.
\newblock {Evolution of the Sun's Polar Fields during Sunspot Cycle 21:
  Poleward Surges and Long-Term Behavior}.
\newblock \emph{\apj}, 347:\penalty0 529, December 1989{\natexlab{b}}.
\newblock \doi{10.1086/168143}.

\bibitem[{Baumann} et~al.(2004){Baumann}, {Schmitt}, {Sch{\"u}ssler}, and
  {Solanki}]{Baumann04}
I.~{Baumann}, D.~{Schmitt}, M.~{Sch{\"u}ssler}, and S.~K. {Solanki}.
\newblock {Evolution of the large-scale magnetic field on the solar surface: A
  parameter study}.
\newblock \emph{\aap}, 426:\penalty0 1075--1091, November 2004.
\newblock \doi{10.1051/0004-6361:20048024}.

\bibitem[{Jiang} et~al.(2014{\natexlab{a}}){Jiang}, {Cameron}, and
  {Sch{\"u}ssler}]{Jiang14}
J.~{Jiang}, R.~H. {Cameron}, and M.~{Sch{\"u}ssler}.
\newblock {Effects of the Scatter in Sunspot Group Tilt Angles on the
  Large-scale Magnetic Field at the Solar Surface}.
\newblock \emph{\apj}, 791:\penalty0 5, August 2014{\natexlab{a}}.
\newblock \doi{10.1088/0004-637X/791/1/5}.

\bibitem[{Thompson} et~al.(1996){Thompson}, {Toomre}, {Anderson}, {Antia},
  {Berthomieu}, {Burtonclay}, {Chitre}, {Christensen-Dalsgaard}, {Corbard}, {De
  Rosa}, {Genovese}, {Gough}, {Haber}, {Harvey}, {Hill}, {Howe}, {Korzennik},
  {Kosovichev}, {Leibacher}, {Pijpers}, {Provost}, {Rhodes}, {Schou}, {Sekii},
  {Stark}, and {Wilson}]{Thompson96}
M.~J. {Thompson}, J.~{Toomre}, E.~R. {Anderson}, H.~M. {Antia},
  G.~{Berthomieu}, D.~{Burtonclay}, S.~M. {Chitre}, J.~{Christensen-Dalsgaard},
  T.~{Corbard}, M.~{De Rosa}, C.~R. {Genovese}, D.~O. {Gough}, D.~A. {Haber},
  J.~W. {Harvey}, F.~{Hill}, R.~{Howe}, S.~G. {Korzennik}, A.~G. {Kosovichev},
  J.~W. {Leibacher}, F.~P. {Pijpers}, J.~{Provost}, E.~J. {Rhodes}, Jr.,
  J.~{Schou}, T.~{Sekii}, P.~B. {Stark}, and P.~R. {Wilson}.
\newblock {Differential Rotation and Dynamics of the Solar Interior}.
\newblock \emph{Science}, 272:\penalty0 1300--1305, May 1996.
\newblock \doi{10.1126/science.272.5266.1300}.

\bibitem[{Antia} et~al.(2008){Antia}, {Basu}, and {Chitre}]{ABC08}
H.~M. {Antia}, S.~{Basu}, and S.~M. {Chitre}.
\newblock {Solar Rotation Rate and Its Gradients During Cycle 23}.
\newblock \emph{\apj}, 681:\penalty0 680--692, July 2008.
\newblock \doi{10.1086/588523}.

\bibitem[{Schou} et~al.(1998){Schou}, {Antia}, {Basu}, {Bogart}, {Bush},
  {Chitre}, {Christensen-Dalsgaard}, {Di Mauro}, {Dziembowski}, {Eff-Darwich},
  {Gough}, {Haber}, {Hoeksema}, {Howe}, {Korzennik}, {Kosovichev}, {Larsen},
  {Pijpers}, {Scherrer}, {Sekii}, {Tarbell}, {Title}, {Thompson}, and
  {Toomre}]{Schou98}
J.~{Schou}, H.~M. {Antia}, S.~{Basu}, R.~S. {Bogart}, R.~I. {Bush}, S.~M.
  {Chitre}, J.~{Christensen-Dalsgaard}, M.~P. {Di Mauro}, W.~A. {Dziembowski},
  A.~{Eff-Darwich}, D.~O. {Gough}, D.~A. {Haber}, J.~T. {Hoeksema}, R.~{Howe},
  S.~G. {Korzennik}, A.~G. {Kosovichev}, R.~M. {Larsen}, F.~P. {Pijpers}, P.~H.
  {Scherrer}, T.~{Sekii}, T.~D. {Tarbell}, A.~M. {Title}, M.~J. {Thompson}, and
  J.~{Toomre}.
\newblock {Helioseismic Studies of Differential Rotation in the Solar Envelope
  by the Solar Oscillations Investigation Using the Michelson Doppler Imager}.
\newblock \emph{\apj}, 505:\penalty0 390--417, September 1998.
\newblock \doi{10.1086/306146}.

\bibitem[{Antia} et~al.(1998){Antia}, {Basu}, and {Chitre}]{ABC98}
H.~M. {Antia}, Sarbani {Basu}, and S.~M. {Chitre}.
\newblock {Solar internal rotation rate and the latitudinal variation of the
  tachocline}.
\newblock \emph{\mnras}, 298\penalty0 (2):\penalty0 543--556, Aug 1998.
\newblock \doi{10.1046/j.1365-8711.1998.01635.x}.

\bibitem[{Gizon} et~al.(2020){Gizon}, {Cameron}, {Pourabdian}, {Liang},
  {Fournier}, {Birch}, and {Hanson}]{Gizon2020}
Laurent {Gizon}, Robert~H. {Cameron}, Majid {Pourabdian}, Zhi-Chao {Liang},
  Damien {Fournier}, Aaron~C. {Birch}, and Chris~S. {Hanson}.
\newblock {Meridional flow in the Sun{\textquoteright}s convection zone is a
  single cell in each hemisphere}.
\newblock \emph{Science}, 368\penalty0 (6498):\penalty0 1469--1472, June 2020.
\newblock \doi{10.1126/science.aaz7119}.

\bibitem[{Passos} et~al.(2015){Passos}, {Charbonneau}, and {Miesch}]{Passos15}
D{\'a}rio {Passos}, Paul {Charbonneau}, and Mark {Miesch}.
\newblock {Meridional Circulation Dynamics from 3D Magnetohydrodynamic Global
  Simulations of Solar Convection}.
\newblock \emph{\apjl}, 800\penalty0 (1):\penalty0 L18, Feb 2015.
\newblock \doi{10.1088/2041-8205/800/1/L18}.

\bibitem[{Hazra} et~al.(2017){Hazra}, {Choudhuri}, and {Miesch}]{HCM17}
G.~{Hazra}, A.~R. {Choudhuri}, and M.~S. {Miesch}.
\newblock {A Theoretical Study of the Build-up of the Sun's Polar Magnetic
  Field by using a 3D Kinematic Dynamo Model}.
\newblock \emph{\apj}, 835:\penalty0 39, January 2017.
\newblock \doi{10.3847/1538-4357/835/1/39}.

\bibitem[{Yoshimura}(1975)]{Yoshimura75}
H.~{Yoshimura}.
\newblock {Solar-cycle dynamo wave propagation}.
\newblock \emph{\apj}, 201:\penalty0 740--748, November 1975.
\newblock \doi{10.1086/153940}.

\bibitem[Choudhuri and Hazra(2016)]{CH16}
Arnab~Rai Choudhuri and Gopal Hazra.
\newblock The treatment of magnetic buoyancy in flux transport dynamo models.
\newblock \emph{Advances in Space Research}, 2016.

\bibitem[{Mu{\~n}oz-Jaramillo} et~al.(2010){Mu{\~n}oz-Jaramillo}, {Nandy},
  {Martens}, and {Yeates}]{Munoz10}
A.~{Mu{\~n}oz-Jaramillo}, D.~{Nandy}, P.~C.~H. {Martens}, and A.~R. {Yeates}.
\newblock {A Double-ring Algorithm for Modeling Solar Active Regions: Unifying
  Kinematic Dynamo Models and Surface Flux-transport Simulations}.
\newblock \emph{\apjl}, 720:\penalty0 L20--L25, September 2010.
\newblock \doi{10.1088/2041-8205/720/1/L20}.

\bibitem[{Yeates} and {Mu{\~n}oz-Jaramillo}(2013)]{YM13}
A.~R. {Yeates} and A.~{Mu{\~n}oz-Jaramillo}.
\newblock {Kinematic active region formation in a three-dimensional solar
  dynamo model}.
\newblock \emph{\mnras}, 436:\penalty0 3366--3379, December 2013.
\newblock \doi{10.1093/mnras/stt1818}.

\bibitem[{Miesch} and {Dikpati}(2014)]{MD14}
M.~S. {Miesch} and M.~{Dikpati}.
\newblock {A Three-dimensional Babcock-Leighton Solar Dynamo Model}.
\newblock \emph{\apjl}, 785:\penalty0 L8, April 2014.
\newblock \doi{10.1088/2041-8205/785/1/L8}.

\bibitem[Miesch and Teweldebirhan(2016)]{MT16}
M.~S. Miesch and K.~Teweldebirhan.
\newblock A three-dimensional babcock-leighton solar dynamo model: Initial
  results with axisymmetric flows.
\newblock \emph{Adv.\ Space Res.}, 58:\penalty0 1571--1588, 2016.

\bibitem[{Karak} and {Miesch}(2018)]{KM18}
B.~B. {Karak} and M.~{Miesch}.
\newblock {Recovery from Maunder-like Grand Minima in a Babcock-Leighton Solar
  Dynamo Model}.
\newblock \emph{\apjl}, 860:\penalty0 L26, June 2018.
\newblock \doi{10.3847/2041-8213/aaca97}.

\bibitem[{Hazra} and {Miesch}(2018)]{HM18}
G.~{Hazra} and M.~S. {Miesch}.
\newblock {Incorporating Surface Convection into a 3D Babcock-Leighton Solar
  Dynamo Model}.
\newblock \emph{\apj}, 864:\penalty0 110, September 2018.
\newblock \doi{10.3847/1538-4357/aad556}.

\bibitem[{Dikpati} and {Choudhuri}(1994)]{DC94}
M.~{Dikpati} and A.~R. {Choudhuri}.
\newblock {The evolution of the Sun's poloidal field}.
\newblock \emph{\aap}, 291:\penalty0 975--989, November 1994.

\bibitem[{Dikpati} and {Choudhuri}(1995)]{DC95}
M.~{Dikpati} and A.~R. {Choudhuri}.
\newblock {On the Large-Scale Diffuse Magnetic Field of the Sun}.
\newblock \emph{\solphys}, 161:\penalty0 9--27, October 1995.
\newblock \doi{10.1007/BF00732081}.

\bibitem[{Choudhuri} and {Dikpati}(1999)]{CD99}
A.~R. {Choudhuri} and M.~{Dikpati}.
\newblock {On the large-scale diffuse magnetic field of the Sun - II.The
  Contribution of Active Regions}.
\newblock \emph{\solphys}, 184:\penalty0 61--76, January 1999.
\newblock \doi{10.1023/A:1005092601436}.

\bibitem[{Baumann} et~al.(2006){Baumann}, {Schmitt}, and
  {Sch{\"u}ssler}]{Baumann06}
I.~{Baumann}, D.~{Schmitt}, and M.~{Sch{\"u}ssler}.
\newblock {A necessary extension of the surface flux transport model}.
\newblock \emph{\aap}, 446:\penalty0 307--314, January 2006.
\newblock \doi{10.1051/0004-6361:20053488}.

\bibitem[{Kumar} et~al.(2019){Kumar}, {Jouve}, and {Nandy}]{Kumar19}
Rohit {Kumar}, Laur{\`e}ne {Jouve}, and Dibyendu {Nandy}.
\newblock {A 3D kinematic Babcock Leighton solar dynamo model sustained by
  dynamic magnetic buoyancy and flux transport processes}.
\newblock \emph{\aap}, 623:\penalty0 A54, Mar 2019.
\newblock \doi{10.1051/0004-6361/201834705}.

\bibitem[{Antia} and {Basu}(2000)]{AB00}
H.~M. {Antia} and Sarbani {Basu}.
\newblock {Temporal Variations of the Rotation Rate in the Solar Interior}.
\newblock \emph{\apj}, 541\penalty0 (1):\penalty0 442--448, Sep 2000.
\newblock \doi{10.1086/309421}.

\bibitem[{Chakraborty} et~al.(2009){Chakraborty}, {Choudhuri}, and
  {Chatterjee}]{CCC09}
S.~{Chakraborty}, A.~R. {Choudhuri}, and P.~{Chatterjee}.
\newblock {Why Does the Sun's Torsional Oscillation Begin before the Sunspot
  Cycle?}
\newblock \emph{{Physical Review Letters}}, 102\penalty0 (4):\penalty0 041102,
  January 2009.
\newblock \doi{10.1103/PhysRevLett.102.041102}.

\bibitem[{Karak} and {Miesch}(2017)]{KM17}
B.~B. {Karak} and M.~{Miesch}.
\newblock {Solar Cycle Variability Induced by Tilt Angle Scatter in a
  Babcock-Leighton Solar Dynamo Model}.
\newblock \emph{\apj}, 847:\penalty0 69, September 2017.
\newblock \doi{10.3847/1538-4357/aa8636}.

\bibitem[{Karak} and {Cameron}(2016)]{KC16}
B.~B. {Karak} and R.~{Cameron}.
\newblock {Babcock-Leighton Solar Dynamo: The Role of Downward Pumping and the
  Equatorward Propagation of Activity}.
\newblock \emph{\apj}, 832:\penalty0 94, November 2016.
\newblock \doi{10.3847/0004-637X/832/1/94}.

\bibitem[{Longcope} and {Choudhuri}(2002)]{Longcope02}
D.~{Longcope} and A.~R. {Choudhuri}.
\newblock {The Orientational Relaxation of Bipolar Active Regions}.
\newblock \emph{\solphys}, 205:\penalty0 63--92, January 2002.
\newblock \doi{10.1023/A:1013896013842}.

\bibitem[{Weber} et~al.(2011){Weber}, {Fan}, and {Miesch}]{Weber11}
Maria~A. {Weber}, Yuhong {Fan}, and Mark~S. {Miesch}.
\newblock {The Rise of Active Region Flux Tubes in the Turbulent Solar
  Convective Envelope}.
\newblock \emph{\apj}, 741\penalty0 (1):\penalty0 11, Nov 2011.
\newblock \doi{10.1088/0004-637X/741/1/11}.

\bibitem[{Stenflo} and {Kosovichev}(2012)]{SK12}
J.~O. {Stenflo} and A.~G. {Kosovichev}.
\newblock {Bipolar Magnetic Regions on the Sun: Global Analysis of the SOHO/MDI
  Data Set}.
\newblock \emph{\apj}, 745:\penalty0 129, February 2012.
\newblock \doi{10.1088/0004-637X/745/2/129}.

\bibitem[{Jiang} et~al.(2015){Jiang}, {Cameron}, and {Sch{\"u}ssler}]{Jiang15}
J.~{Jiang}, R.~H. {Cameron}, and M.~{Sch{\"u}ssler}.
\newblock {The Cause of the Weak Solar Cycle 24}.
\newblock \emph{\apjl}, 808:\penalty0 L28, July 2015.
\newblock \doi{10.1088/2041-8205/808/1/L28}.

\bibitem[{Nagy} et~al.(2017){Nagy}, {Lemerle}, {Labonville}, {Petrovay}, and
  {Charbonneau}]{Nagy17}
Melinda {Nagy}, Alexandre {Lemerle}, Fran{\c{c}}ois {Labonville}, Krist{\'o}f
  {Petrovay}, and Paul {Charbonneau}.
\newblock {The Effect of ``Rogue'' Active Regions on the Solar Cycle}.
\newblock \emph{\solphys}, 292\penalty0 (11):\penalty0 167, Nov 2017.
\newblock \doi{10.1007/s11207-017-1194-0}.

\bibitem[{Kitchatinov} et~al.(1994){Kitchatinov}, {Pipin}, and
  {Ruediger}]{Kitchatinov94}
L.~L. {Kitchatinov}, V.~V. {Pipin}, and G.~{Ruediger}.
\newblock {Turbulent viscosity, magnetic diffusivity, and heat conductivity
  under the influence of rotation and magnetic field}.
\newblock \emph{Astronomische Nachrichten}, 315:\penalty0 157--170, February
  1994.
\newblock \doi{10.1002/asna.2103150205}.

\bibitem[{Charbonneau} and {Dikpati}(2000)]{CD2000}
P.~{Charbonneau} and M.~{Dikpati}.
\newblock {Stochastic Fluctuations in a Babcock-Leighton Model of the Solar
  Cycle}.
\newblock \emph{\apj}, 543:\penalty0 1027--1043, November 2000.
\newblock \doi{10.1086/317142}.

\bibitem[{Karak} and {Choudhuri}(2011)]{KarakChou11}
B.~B. {Karak} and A.~R. {Choudhuri}.
\newblock {The Waldmeier effect and the flux transport solar dynamo}.
\newblock \emph{\mnras}, 410:\penalty0 1503--1512, January 2011.
\newblock \doi{10.1111/j.1365-2966.2010.17531.x}.

\bibitem[{Karak} and {Choudhuri}(2013)]{KarakChou13}
B.~B. {Karak} and A.~R. {Choudhuri}.
\newblock {Studies of grand minima in sunspot cycles by using a flux transport
  solar dynamo model}.
\newblock \emph{Research in Astronomy and Astrophysics}, 13:\penalty0 1339,
  November 2013.
\newblock \doi{10.1088/1674-4527/13/11/005}.

\bibitem[{Hazra} et~al.(2015){Hazra}, {Karak}, {Banerjee}, and
  {Choudhuri}]{HKBC15}
G.~{Hazra}, B.~B. {Karak}, D.~{Banerjee}, and A.~R. {Choudhuri}.
\newblock {Correlation Between Decay Rate and Amplitude of Solar Cycles as
  Revealed from Observations and Dynamo Theory}.
\newblock \emph{\solphys}, 290:\penalty0 1851--1870, June 2015.
\newblock \doi{10.1007/s11207-015-0718-8}.

\bibitem[{Hazra} and {Choudhuri}(2019)]{HC19}
Gopal {Hazra} and Arnab~Rai {Choudhuri}.
\newblock {A New Formula for Predicting Solar Cycles}.
\newblock \emph{\apj}, 880\penalty0 (2):\penalty0 113, Aug 2019.
\newblock \doi{10.3847/1538-4357/ab2718}.

\bibitem[{Jiang} et~al.(2007){Jiang}, {Chatterjee}, and {Choudhuri}]{Jiang07}
J.~{Jiang}, P.~{Chatterjee}, and A.~R. {Choudhuri}.
\newblock {Solar activity forecast with a dynamo model}.
\newblock \emph{\mnras}, 381:\penalty0 1527--1542, November 2007.
\newblock \doi{10.1111/j.1365-2966.2007.12267.x}.

\bibitem[{Wang} et~al.(2009){Wang}, {Robbrecht}, and {Sheeley}]{Wang09}
Y.~M. {Wang}, E.~{Robbrecht}, and Jr. {Sheeley}, N.~R.
\newblock {On the Weakening of the Polar Magnetic Fields during Solar Cycle
  23}.
\newblock \emph{\apj}, 707\penalty0 (2):\penalty0 1372--1386, Dec 2009.
\newblock \doi{10.1088/0004-637X/707/2/1372}.

\bibitem[{Svalgaard} et~al.(2005){Svalgaard}, {Cliver}, and {Kamide}]{SCK05}
L.~{Svalgaard}, E.~W. {Cliver}, and Y.~{Kamide}.
\newblock {Sunspot cycle 24: Smallest cycle in 100 years?}
\newblock \emph{\grl}, 32:\penalty0 L01104, January 2005.
\newblock \doi{10.1029/2004GL021664}.

\bibitem[{Mu{\~n}oz-Jaramillo} et~al.(2013){Mu{\~n}oz-Jaramillo},
  {Dasi-Espuig}, {Balmaceda}, and {DeLuca}]{Munoz13}
Andr{\'e}s {Mu{\~n}oz-Jaramillo}, Mar{\'\i}a {Dasi-Espuig}, Laura~A.
  {Balmaceda}, and Edward~E. {DeLuca}.
\newblock {Solar Cycle Propagation, Memory, and Prediction: Insights from a
  Century of Magnetic Proxies}.
\newblock \emph{\apjl}, 767\penalty0 (2):\penalty0 L25, Apr 2013.
\newblock \doi{10.1088/2041-8205/767/2/L25}.

\bibitem[{Priyal} et~al.(2014){Priyal}, {Banerjee}, {Karak},
  {Mu{\~n}oz-Jaramillo}, {Ravindra}, {Choudhuri}, and {Singh}]{Priyal14}
M.~{Priyal}, D.~{Banerjee}, B.~B. {Karak}, A.~{Mu{\~n}oz-Jaramillo},
  B.~{Ravindra}, A.~R. {Choudhuri}, and J.~{Singh}.
\newblock {Polar Network Index as a Magnetic Proxy for the Solar Cycle
  Studies}.
\newblock \emph{\apjl}, 793:\penalty0 L4, September 2014.
\newblock \doi{10.1088/2041-8205/793/1/L4}.

\bibitem[{Choudhuri}(1992)]{Chou92}
A.~R. {Choudhuri}.
\newblock {Stochastic fluctuations of the solar dynamo}.
\newblock \emph{\aap}, 253:\penalty0 277--285, January 1992.

\bibitem[{Choudhuri} et~al.(2007){Choudhuri}, {Chatterjee}, and {Jiang}]{CCJ07}
A.~R. {Choudhuri}, P.~{Chatterjee}, and J.~{Jiang}.
\newblock {Predicting Solar Cycle 24 With a Solar Dynamo Model}.
\newblock \emph{Physical Review Letters}, 98:\penalty0 131103, March 2007.
\newblock \doi{10.1103/PhysRevLett.98.131103}.

\bibitem[{Choudhuri} and {Karak}(2009)]{CK09}
A.~R. {Choudhuri} and B.~B. {Karak}.
\newblock {A possible explanation of the Maunder minimum from a flux transport
  dynamo model}.
\newblock \emph{Research in Astronomy and Astrophysics}, 9:\penalty0 953--958,
  September 2009.
\newblock \doi{10.1088/1674-4527/9/9/001}.

\bibitem[{Jiang} et~al.(2014{\natexlab{b}}){Jiang}, {Hathaway}, {Cameron},
  {Solanki}, {Gizon}, and {Upton}]{Jiang_review15}
J.~{Jiang}, D.~H. {Hathaway}, R.~H. {Cameron}, S.~K. {Solanki}, L.~{Gizon}, and
  L.~{Upton}.
\newblock {Magnetic Flux Transport at the Solar Surface}.
\newblock \emph{\ssr}, 186:\penalty0 491--523, December 2014{\natexlab{b}}.
\newblock \doi{10.1007/s11214-014-0083-1}.

\bibitem[{Lemerle} and {Charbonneau}(2017{\natexlab{a}})]{Lemerle17}
Alexandre {Lemerle} and Paul {Charbonneau}.
\newblock {A Coupled 2 {$\times$} 2D Babcock-Leighton Solar Dynamo Model. II.
  Reference Dynamo Solutions}.
\newblock \emph{\apj}, 834\penalty0 (2):\penalty0 133, Jan 2017{\natexlab{a}}.
\newblock \doi{10.3847/1538-4357/834/2/133}.

\bibitem[{Howard}(1991)]{Howard91}
Robert~F. {Howard}.
\newblock {Axial Tilt Angles of Sunspot Groups}.
\newblock \emph{\solphys}, 136\penalty0 (2):\penalty0 251--262, Dec 1991.
\newblock \doi{10.1007/BF00146534}.

\bibitem[{Wang} et~al.(2015){Wang}, {Colaninno}, {Baranyi}, and {Li}]{Wang15}
Y.-M. {Wang}, R.~C. {Colaninno}, T.~{Baranyi}, and J.~{Li}.
\newblock {Active-region Tilt Angles: Magnetic versus White-light
  Determinations of Joy's Law}.
\newblock \emph{\apj}, 798:\penalty0 50, January 2015.
\newblock \doi{10.1088/0004-637X/798/1/50}.

\bibitem[{Chatterjee} and {Choudhuri}(2006)]{CC06}
P.~{Chatterjee} and A.~R. {Choudhuri}.
\newblock {On Magnetic Coupling Between the Two Hemispheres in Solar Dynamo
  Models}.
\newblock \emph{\solphys}, 239:\penalty0 29--39, December 2006.
\newblock \doi{10.1007/s11207-006-0201-6}.

\bibitem[{Lemerle} and {Charbonneau}(2017{\natexlab{b}})]{Lemerle15}
Alexandre {Lemerle} and Paul {Charbonneau}.
\newblock {A Coupled 2 {$\times$} 2D Babcock-Leighton Solar Dynamo Model. II.
  Reference Dynamo Solutions}.
\newblock \emph{\apj}, 834\penalty0 (2):\penalty0 133, Jan 2017{\natexlab{b}}.
\newblock \doi{10.3847/1538-4357/834/2/133}.

\bibitem[{Hathaway}(2012{\natexlab{a}})]{Hathaway12a}
D.~H. {Hathaway}.
\newblock {Supergranules as Probes of Solar Convection Zone Dynamics}.
\newblock \emph{\apjl}, 749:\penalty0 L13, April 2012{\natexlab{a}}.
\newblock \doi{10.1088/2041-8205/749/1/L13}.

\bibitem[{Hathaway}(2012{\natexlab{b}})]{Hathaway12}
D.~H. {Hathaway}.
\newblock {Supergranules as Probes of the Sun's Meridional Circulation}.
\newblock \emph{\apj}, 760:\penalty0 84, November 2012{\natexlab{b}}.
\newblock \doi{10.1088/0004-637X/760/1/84}.

\bibitem[{Charbonneau}(2014)]{Charbonneau14}
P.~{Charbonneau}.
\newblock {Solar Dynamo Theory}.
\newblock \emph{\araa}, 52:\penalty0 251--290, August 2014.
\newblock \doi{10.1146/annurev-astro-081913-040012}.

\bibitem[{Rempel}(2005)]{Rempel05}
M.~{Rempel}.
\newblock {Solar Differential Rotation and Meridional Flow: The Role of a
  Subadiabatic Tachocline for the Taylor-Proudman Balance}.
\newblock \emph{\apj}, 622:\penalty0 1320--1332, April 2005.
\newblock \doi{10.1086/428282}.

\bibitem[{Whitbread} et~al.(2019){Whitbread}, {Yeates}, and
  {Mu{\~n}oz-Jaramillo}]{whitbread19}
T.~{Whitbread}, A.~R. {Yeates}, and A.~{Mu{\~n}oz-Jaramillo}.
\newblock {The need for active region disconnection in 3D kinematic dynamo
  simulations}.
\newblock \emph{\aap}, 627:\penalty0 A168, Jul 2019.
\newblock \doi{10.1051/0004-6361/201935986}.

\end{thebibliography}


\end{document}